\PassOptionsToPackage{longnamesfirst,sort}{natbib}
\documentclass[aps,reprint,superscriptaddress,notitlepage,floatfix]{revtex4-2}
\usepackage{layout-compat}
\usepackage{booktabs}

\usepackage{amsmath,amssymb,bm}
\usepackage{booktabs}
\usepackage{graphicx}

\newcommand{\NpT}{N_{p,T}}
\newcommand{\WT}{W_T}
\newcommand{\hT}{h_T}
\newcommand{\lambar}{\overline{\lambda}}
\newcommand{\E}{\mathbb{E}}
\newcommand{\Prob}{\mathbb{P}}

\begin{document}

\title{Finite-Horizon Triggering Cores\\ of Earthquake Sequences}

\author{Didier Sornette}
\affiliation{Institute of Risk Analysis, Prediction and Management (Risks-X), Academy for Advanced Interdisciplinary Studies, Southern University of Science and Technology, Shenzhen, China}
\author{Giuseppe Petrillo}
\affiliation{CNRS, ENS de Lyon, LPENSL, UMR 5672, 69342 Lyon, France}

\authoraddr{Giuseppe Petrillo, Laboratoire de Physique, ENS de Lyon, 46 all\'ee d'Italie, 69364 Lyon Cedex 07, France (giuseppe51289@gmail.com)}

\authorrunninghead{Ignored.}
\titlerunninghead{Ignored.}

\begin{abstract}
Models of triggered seismicity commonly aggregate the contributions of all past earthquakes into a total predicted rate. This aggregation does not distinguish a forecast dominated by one recent earthquake from an equal-rate forecast supported by many events distributed across a longer history, even though these configurations can have different persistence and sensitivity to catalog and parameter uncertainties. We introduce the \emph{finite-horizon triggering core size} \(N_{p,T}(t)\), defined as the minimum number of past earthquakes whose expected contributions account for a fraction \(p\) of the total triggering potential over the future interval \([t,t+T]\). The definition applies to any triggering model that assigns nonnegative event-specific expected contributions over a specified future window. We develop its theory for marked Hawkes and epidemic-type aftershock sequence processes, where the weights are the expected numbers of direct offspring produced by individual past earthquakes. The finite horizon makes the observable well defined for every normalized Omori kernel with exponent \(\theta>0\), including the long-memory regime in which the corresponding infinite-horizon quantity diverges. We derive continuum and marked-Poisson benchmarks, an exact representation of the marked-Poisson distribution, and high-\(p\) distributional laws. For finite-horizon Omori memory, the triggering core grows as \((1-p)^{-1/\theta}\) with a realization-dependent amplitude controlled by the total triggering weight; for exponential memory, the growth is logarithmic. Exact simulations validate these predictions and show that the leading marked-Poisson scaling remains accurate for the tested self-exciting histories. Transient experiments reveal continuous redistribution and rank crossings under Omori memory, whereas exponential normalized weights change only at event arrivals. Calendar-time occupation laws self-average increasingly slowly near critical branching, while event-time sampling converges to a distinct Palm law. Together, the predicted rate and \(N_{p,T}(t)\) separate the level of expected seismicity from the concentration of its triggering sources, providing a general diagnostic of forecast attribution, persistence, and sensitivity.

\end{abstract}

\maketitle

\section{Introduction}

Self-exciting point processes provide a parsimonious description of systems in which each event
temporarily increases the probability of later events. The foundational Hawkes formulation and
its Poisson-cluster representation made the connection between conditional intensity and
branching explicit \cite{Hawkes1971,HawkesOakes1974}; a broad review of spatiotemporal
self-exciting processes is given by \citeA{Reinhart2018}. In seismology, stochastic branching
catalogs were developed by \citeA{KaganKnopoff1981}, and the epidemic-type aftershock sequence
(ETAS) model placed this idea in a likelihood-based point-process framework
\cite{Ogata1988,Ogata1998}. ETAS treats background earthquakes as immigrants and every earthquake
as a possible parent of direct aftershocks, which can trigger further generations. Aftershocks, foreshocks, 
and mainshocks are therefore not distinct event types in the model; rather, they are history-dependent roles assumed by earthquakes within a single triggering cascade
 \cite{HelmstetterSornetteGrasso2003,HelmstetterSornette2003}.

The appeal of ETAS comes in part from its synthesis of robust empirical regularities. Earthquake
magnitudes follow the Gutenberg--Richter law \cite{GutenbergRichter1944}; direct-aftershock rates
decay according to a modified Omori kernel \cite{Utsu1995}; and aftershock productivity grows approximately exponentially with mainshock magnitude \cite{Utsu1970a,Utsu1970b}, a relation incorporated in ETAS as an exponential dependence of the expected number of direct offspring on parent magnitude. These ingredients imply a Pareto distribution
of event fertilities and can generate large realization-to-realization variability in cluster size
and duration \cite{HelmstetterSornette2002,SaichevHelmstetterSornette2005}. The branching ratio 
$n$, defined as the mean number of direct offspring per event, determines the dynamical regime: 
subcritical for $n<1$, critical for $n=1$, and supercritical for $n>1$.
The ratio between the Gutenberg-Richter and productivity exponents
controls fertility heterogeneity. Both the smallest triggering magnitude and undetected events
can materially alter inferred productivity and branching
\cite{SornetteWerner2005a,SornetteWerner2005b,WernerSornette2008}. Moreover,
empirical productivity fluctuates even among mainshocks of similar magnitude
\cite{MarsanHelmstetter2017}. Concentration is consequently a random property of the whole recent
history, not a deterministic function of seismicity rate alone.

The ETAS conditional intensity is the instantaneous earthquake occurrence rate predicted at a given time from the complete prior seismic history, combining the background rate with the contributions of all preceding earthquakes that may trigger subsequent events. The ETAS conditional intensity is indispensable for fitting, forecasting, and residual analysis, but it is an aggregate. At a given time, it sums the background component and the contributions of
all previous earthquakes. Statistical declustering and stochastic reconstruction recover
probabilistic parent assignments or cluster structure from this sum
\cite{Zhuang2002,Zhuang2004,VeenSchoenberg2008,MarsanLengline2008}; complementary nearest-neighbor methods address related clustering and declustering questions without relying on the specific background-versus-triggered decomposition assumed by ETAS \cite{ZaliapinBenZion2020,bountzis2026automatic}. These methods ask which earlier event is a plausible
parent of each observed earthquake. The question considered here is different and
forecast-oriented: among all past events, how many currently carry most of the expected direct
triggering over a specified future interval?

This distinction also separates triggering concentration from standard model diagnostics.
Time-rescaling, thinning, superposition, and spatial residual methods test whether a fitted
conditional intensity is compatible with the observed point pattern
\cite{Ogata1988,Schoenberg2003,Clements2011}. Earthquake forecast testing likewise compares
predicted and observed counts or likelihoods in prescribed bins \cite{Schorlemmer2007}. Such tests
can detect a wrong total rate or a wrong space--time distribution, but they do not directly report
whether the endogenous part of a correct total rate is dominated by one parent, a few comparable
parents, or a diffuse population of old events. Triggering concentration is intended to complement,
not replace, these established residual and likelihood diagnostics.

\begin{figure*}[!tp] 
\centering \includegraphics[width=\textwidth,height=0.66\textheight,keepaspectratio] {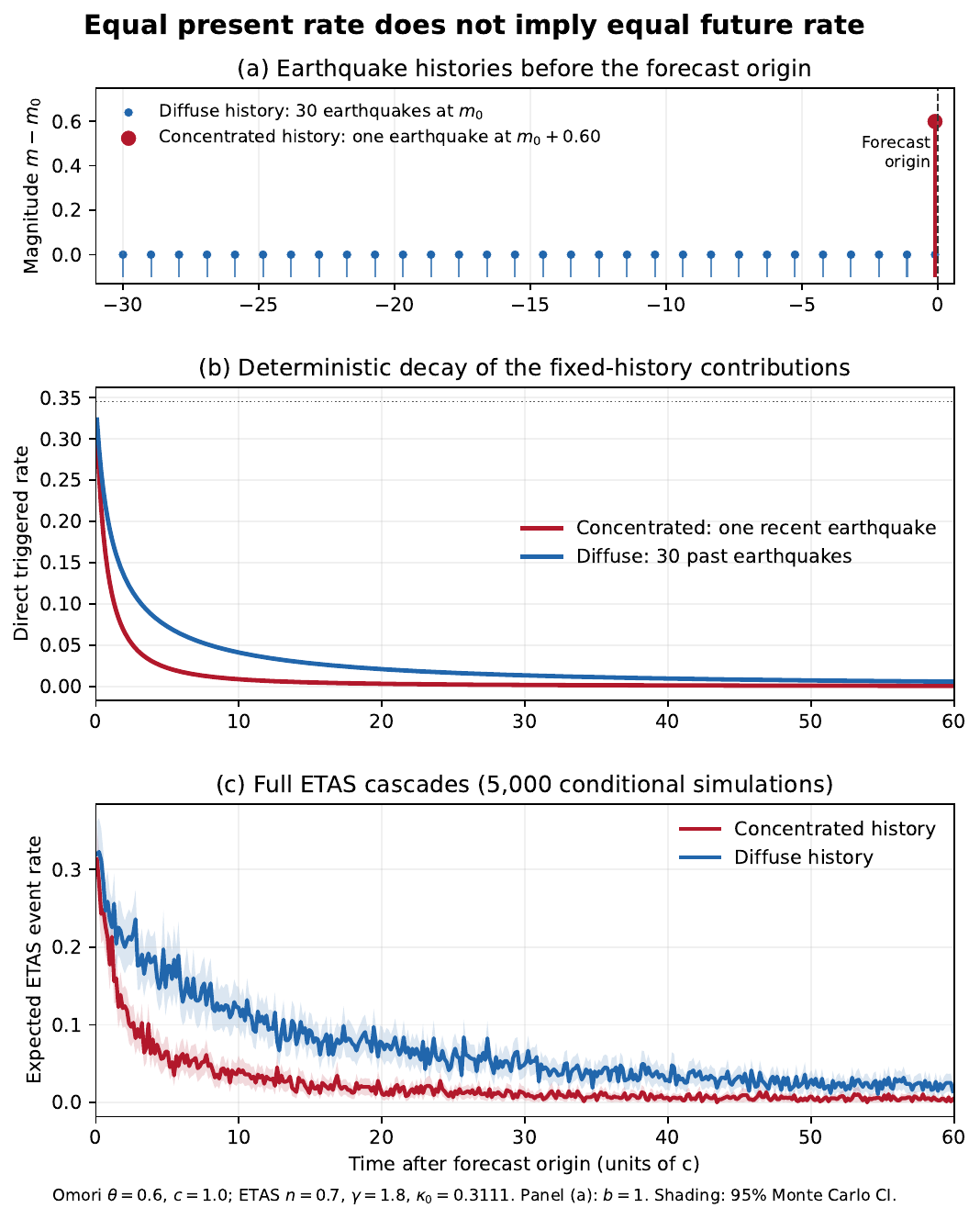} 
\caption{Source histories and subsequent rate evolution for two earthquake histories having the same endogenous rate, \(0.34513\), at the forecast origin \(t=0\). The normalized Omori kernel is \(\phi(\tau)=\theta c^\theta(\tau+c)^{-(1+\theta)}\), with \(\theta=0.6\) and \(c=1\). The ETAS parameters are \(n=0.7\), \(\gamma=1.8\), and \(\kappa_0=n(\gamma-1)/\gamma=0.3111\). The concentrated history contains one earthquake of age \(0.1c\) and fertility \(\kappa_{\mathrm{single}}=0.66998\). The diffuse history contains 30 earthquakes of fertility \(\kappa_0\), with ages uniformly distributed between \(0.1c\) and \(30c\). For the magnitude representation in panel~(a), \(b=1\), so that the concentrated earthquake has magnitude \(m_0+0.60\), while the 30 earthquakes in the diffuse history have magnitude \(m_0\). (a) Occurrence times and magnitudes \(m-m_0\) of the earthquakes preceding the forecast origin. (b) Direct triggered rates generated by the two fixed histories. (c) Mean event rates obtained from 5,000 conditional ETAS simulations including all descendant generations; shaded regions denote 95\% Monte Carlo confidence intervals for the mean.} \label{fig:etas_equal_rate_histories} 
\end{figure*}

Two histories illustrate the information lost through aggregation. One may contain a single recent and relatively productive earthquake that accounts for nearly the entire endogenous intensity. Another may contain many weaker earthquakes distributed over an extended past, whose individual contributions are small but whose sum is identical. The conditional intensities of these histories can therefore coincide at the forecast origin while their subsequent evolutions differ sharply. In the concentrated case, the endogenous rate is governed essentially by the decay of a single memory kernel. In the diffuse case, it is a superposition of kernels initiated at different times and therefore observed at different stages of their decay. As the contributions of the youngest earthquakes weaken, the older components collectively sustain a longer-lived tail. The two forecasts consequently differ in their attribution, sensitivity, and temporal persistence. In the concentrated case, uncertainty in one magnitude, one productivity factor, or the short-term detection record may control most of the forecast. In the diffuse case, no single earthquake is decisive, but the forecast depends on the accurate representation of a longer history. 

Figure~\ref{fig:etas_equal_rate_histories} illustrates this distinction using a temporal ETAS experiment. The concentrated history contains one earthquake of age \(0.1c\) and fertility \(\kappa_{\mathrm{single}}=0.66998\). The diffuse history contains 30 earthquakes of fertility \(\kappa_0=0.3111\), with ages uniformly distributed between \(0.1c\) and \(30c\). The fertility of the single earthquake is chosen so that the two histories generate exactly the same endogenous rate, \(0.34513\), at the forecast origin. For \(b=1\), the corresponding magnitudes are \(m_0+0.60\) for the single earthquake and \(m_0\) for each earthquake in the diffuse history. Despite their identical initial rates, at \(t=30c\) the direct rate generated by the diffuse history is approximately \(8.1\) times that generated by the concentrated history. Over the interval \([0,30c]\), 5,000 conditional ETAS simulations including all descendant generations produce an average of \(3.24\) events from the diffuse history, compared with \(1.26\) from the concentrated history. Equality of the present conditional intensity therefore does not imply equality of either the future rate trajectory or the expected number of future earthquakes. 

To distinguish histories that produce the same present conditional intensity but
differ in the concentration, sensitivity, and temporal persistence of their future
triggering, as illustrated in Figure~\ref{fig:etas_equal_rate_histories}, we
introduce \(N_{p,T}(t)\). In the Hawkes/ETAS setting, \(N_{p,T}(t)\) is the
minimum number of past earthquakes whose expected direct offspring in the future
interval \([t,t+T]\) account for a fraction \(p\) of the total expected direct
triggering over that interval. We call \(N_{p,T}(t)\) the
\emph{finite-horizon triggering core size}.

The concept is not specific to Hawkes or ETAS processes. It can be defined for any
triggering model that decomposes the expected activity over a future interval into
nonnegative contributions attributable to individual past events. Hawkes/ETAS
processes provide the analytical framework developed here and the natural
seismological setting in which each contribution is the expected number of direct
offspring of a past earthquake.

The finite-horizon triggering core size is a ranked-share statistic, related
conceptually to cumulative concentration curves \cite{Lorenz1905}. We complement
the full curve \(p\mapsto N_{p,T}(t)\) with inverse-participation and
entropy-based effective population sizes, which summarize different aspects of the
distribution of normalized triggering weights
\cite{Simpson1949,Hill1973,Jost2006}.

The finite horizon is a mathematical and operational necessity. If an event of age \(a\) has
fertility \(\kappa\), its infinite-future residual direct fertility is \(\kappa\Phi(a)\), where
\(\Phi\) is the survival function of the triggering kernel. For a normalized Omori density
\(\phi(a)\propto(a+c)^{-1-\theta}\), each individual residual is finite, but their sum over an
infinite stationary past has finite mean only when \(\int_0^\infty\Phi(a)\,da<\infty\), namely
\(\theta>1\). The empirically important long-memory range \(0<\theta\le1\) therefore makes the
infinite-forecast concentration of a stationary history ill posed. Replacing \(\Phi(a)\) by the
finite-window mass \(\Phi(a)-\Phi(a+T)\) restores integrability for every \(\theta>0\) and ties the
observable to an explicit forecast horizon rather than to an arbitrary catalog cutoff.

A first-moment or continuum calculation is not sufficient. The threshold that captures fraction
\(p\) of the total triggering weight is itself random and correlated with that total. Heavy-tailed
fertilities further separate typical realizations from ensemble means
\cite{SaichevSornette2004,SaichevSornette2014}. Recent analytical works have derived the full space–time–magnitude distribution of the largest event in a subcritical ETAS cluster \cite{spassiani2024distribution,petrillo2026full}. Here we address a complementary distributional question: rather than the properties of the largest future event, we characterize how the expected direct triggering over a finite future horizon is concentrated among past earthquakes. We therefore study the full law of \(N_{p,T}\).
For a stationary marked-Poisson history, mapping ages and fertilities to weights produces a
one-dimensional Poisson point process and an exact annealed representation. Exponential memory
leads to a scale-invariant weight cloud connected to Poisson--Dirichlet and generalized Dickman
structures \cite{Arratia1999,Pitman2003}; finite-horizon Omori memory instead produces a universal
small-weight power law and a random high-quantile amplitude. Hawkes genealogical clustering then
provides a controlled test of how far the Poisson distributional laws extend beyond independent
histories.

Distributional statements also depend on how time is sampled. The annealed law draws independent
stationary histories at a calendar origin. A single long catalog produces an occupation law along
calendar time, which should converge to the annealed law under ordinary subcritical ergodicity.
Sampling only at earthquake times yields a Palm law and is intensity biased even when the process
is perfectly stationary. Stable Hawkes laws of large numbers support calendar-time self-averaging
under suitable conditions \cite{BremaudMassoulie1996,Bacry2013,Zhu2013}, while nearly unstable and
heavy-memory regimes can converge exceptionally slowly
\cite{JaissonRosenbaum2015,JaissonRosenbaum2016}. Separating these three laws prevents finite-record
dependence or event-time bias from being mislabeled as ergodicity breaking.

The present paper develops the theory and numerical verification layer of this program. We (1)
define the finite-horizon observables and their integrability domain; (2) derive constant-fertility,
ETAS/Pareto, exponential, Omori, and finite-lookback crossover formulas; (3) obtain the marked-
Poisson annealed law and finite-horizon high-\(p\) scaling; (4) characterize the Hawkes threshold law
by a cluster generating functional; and (5) validate the predictions with exact cluster
simulations, including finite-history controls, Monte Carlo uncertainty, near-critical occupation
tests, transient condensation, and rank dynamics.

\section{Finite-Horizon Triggering Cores}

\subsection{General definition}

Consider an event history \(\mathcal H_t\) observed up to time \(t\), and a model that
attributes nonnegative expected future triggering contributions to the individual past
events. Specifically, let
\[
 g_i(u\mid\mathcal H_t)\geq0,
 \qquad u>t,
\]
denote the contribution attributed to past event \(i\) to the expected triggered-event
rate at future time \(u\). The contribution may depend on the properties of event \(i\),
its age, the remaining history, and the parameters of the triggering model. The only
structure required here is that the model supplies an additive eventwise decomposition
of the expected triggered activity.

For a forecast horizon \(T>0\), define the finite-horizon triggering weight of event
\(i\) by
\begin{equation}
 w_{i,T}(t)
 =
 \int_t^{t+T}
 g_i(u\mid\mathcal H_t)\,du.
 \label{eq:general_weight}
\end{equation}
Thus, \(w_{i,T}(t)\) is the expected triggering contribution assigned by the model to
event \(i\) over the future interval \([t,t+T]\). The total finite-horizon triggering
potential carried by the observed history is
\begin{equation}
 W_T(t)
 =
 \sum_{t_i<t}w_{i,T}(t).
 \label{eq:general_total_weight}
\end{equation}
The background or exogenous component is not included in \(W_T(t)\), because the
purpose is to characterize how the endogenous triggering potential is distributed
among past events.

Assume that \(0<W_T(t)<\infty\), and order the event-specific weights as
\[
 w_{(1),T}(t)\geq w_{(2),T}(t)\geq\cdots.
\]
For \(0<p<1\), define
\begin{equation}
 N_{p,T}(t)
 =
 \min\left\{
 k:
 \sum_{j=1}^{k}w_{(j),T}(t)
 \geq pW_T(t)
 \right\}.
 \label{eq:NpT}
\end{equation}
We call \(N_{p,T}(t)\) the \emph{finite-horizon triggering core size}. It is the
smallest number of past events whose combined model-implied contributions account for
at least the fraction \(p\) of the total expected triggering over \([t,t+T]\).
The corresponding events form the finite-horizon triggering core at level \(p\).

A small value of \(N_{p,T}(t)\) identifies a concentrated state in which a few past
events control most of the expected triggering. A large value identifies a diffuse
state in which triggering responsibility is distributed across a broader history. For
example, \(N_{0.9,30\,\mathrm{d}}(t)=4\) means that four past earthquakes account
for at least \(90\%\) of the model-implied triggering potential during the next
30 days. The full curve
\[
 p\longmapsto N_{p,T}(t)
\]
characterizes the cumulative concentration of the ranked triggering contributions.

This definition is not specific to Hawkes or ETAS processes. It applies whenever a
triggering model supplies nonnegative event-specific expected contributions over a
future window. Different models may define these contributions through direct
offspring, cascade-adjusted descendants, stress transfer, space--time influence, or
another explicit attribution rule. The interpretation of the resulting triggering
core must therefore be stated together with the attribution convention used to define
the weights.

For a finite catalog containing \(M\) past events, one may additionally report
\begin{equation}
 f_{p,T}(t)=\frac{N_{p,T}(t)}{M},
\end{equation}
the fraction of cataloged events belonging to the triggering core. This normalization
depends on the catalog duration and completeness. In a stationary construction with
an infinite past and strictly positive weights, infinitely many events contribute to
\(W_T(t)\), so \(f_{p,T}(t)\) has no nontrivial infinite-history limit. By contrast,
\(N_{p,T}(t)\) remains finite for \(p<1\) whenever the ranked weights have a finite
positive sum. The effective population sizes introduced below provide complementary
cutoff-free summaries of the normalized weight distribution.

\subsection{Specialization to marked Hawkes and ETAS processes}

We now specialize the general construction to a scalar marked Hawkes process. Its
conditional intensity is
\begin{equation}
 \lambda(t)
 =
 \nu+\sum_{t_i<t}\kappa_i\phi(t-t_i),
 \qquad
 \int_0^\infty\phi(s)\,ds=1,
 \label{eq:intensity}
\end{equation}
where \(\nu>0\) is the immigrant or background rate,
\(\kappa_i\geq0\) is the fertility mark of earthquake \(i\), and \(\phi\) is a
normalized causal triggering kernel. Unless stated otherwise, the fertility marks are
independent and identically distributed and are independent of event times and
genealogy.

Each earthquake produces a Poisson number of direct offspring with conditional mean
\(\kappa_i\). The branching ratio is therefore
\begin{equation}
 n=\E[\kappa].
\end{equation}
For \(n<1\), the standard stationary linear Hawkes process has finite mean rate
\begin{equation}
 \lambar=\frac{\nu}{1-n}
\end{equation}
under the usual stability assumptions \cite{BremaudMassoulie1996}. This subcritical
stationarity condition is distinct from scaling crossovers involving the Omori and
fertility exponents.

Define the survival function of the triggering kernel and its finite-window mass by
\begin{equation}
\begin{gathered}
 \Phi(a)
 =
 \int_a^\infty\phi(s)\,ds,
 \\ \qquad
 h_T(a)
 =
 \Phi(a)-\Phi(a+T)
 =
 \int_a^{a+T}\phi(s)\,ds.
 \label{eq:horizon_kernel}
\end{gathered}
\end{equation}
At evaluation time \(t\), earthquake \(i\) has age \(a_i=t-t_i>0\). Its
contribution to the expected direct triggering rate at a future time \(u>t\) is
\[
 g_i(u\mid\mathcal H_t)
 =
 \kappa_i\phi(u-t_i).
\]
Substitution into the general definition
\eqref{eq:general_weight} gives
\begin{equation}
 w_{i,T}(t)
 =
 \int_t^{t+T}
 \kappa_i\phi(u-t_i)\,du
 =
 \kappa_i h_T(a_i).
 \label{eq:weight}
\end{equation}
Hence, in the Hawkes/ETAS application, \(w_{i,T}(t)\) is the conditional expected
number of \emph{direct} offspring of earthquake \(i\) in \([t,t+T]\). It excludes
descendants mediated by future offspring. This convention makes the attribution
eventwise and allows the weights and the triggering core to be computed without
reconstructing the realized parent--offspring genealogy or simulating future catalogs.

For a given Hawkes/ETAS history
\(\{(t_i,\kappa_i):t_i<t\}\), the total direct triggering potential is therefore
\begin{equation}
 W_T(t)
 =
 \sum_{t_i<t}
 \kappa_i h_T(t-t_i),
 \label{eq:hawkes_total_weight}
\end{equation}
and its finite-horizon triggering core size is obtained by ranking these contributions
and applying equation~\eqref{eq:NpT}. All subsequent analytical results in this paper
use this direct-offspring Hawkes/ETAS specialization of the general triggering-core
concept.

\subsection{Integrability and the role of the finite horizon}

By Tonelli's theorem, stated for example as Theorem~2.37 in
\cite{Folland1999}, the nonnegativity of the integrand permits the order of
integration to be exchanged, yielding
\begin{equation}
 \int_0^\infty\hT(a)\,da
 =\int_0^T\Phi(u)\,du\le T.
 \label{eq:tonelli}
\end{equation}
Consequently,
\begin{equation}
 \E[\WT]=\lambar n\int_0^T\Phi(u)\,du<\infty,
 \label{eq:meanW}
\end{equation}
and \(\WT<\infty\) almost surely in a stationary finite-rate process.

For the normalized Omori kernel
\begin{equation}
 \phi(a)=\frac{\theta c^\theta}{(a+c)^{1+\theta}},
 \qquad
 \Phi(a)=\left(1+\frac{a}{c}\right)^{-\theta},
 \label{eq:omori}
\end{equation}
the infinite-future mean residual activity involves \(\int_0^\infty\Phi(a)da\), which is finite
only for \(\theta>1\). In contrast, equation~\eqref{eq:tonelli} is finite for every \(\theta>0\).
The divergence at \(T=\infty\) for \(\theta\le1\) is not a failure of Hawkes stationarity:
\(\phi\) remains normalized, and ordinary stationarity still requires \(n<1\).

An exponential kernel with time scale \(c=1/\omega\) has
\begin{equation}
 \phi(a)=\omega e^{-\omega a},\qquad
 \hT(a)=(1-e^{-\omega T})e^{-\omega a}.
 \label{eq:exponential} 
\end{equation}
The horizon-dependent factor is common to all events, so rankings and normalized shares are
exactly independent of \(T\). Omori \(\hT\) does not factorize this way; changing \(T\) can change
ranks, and relative weights can cross continuously between event arrivals.

\subsection{Evaluation from an observed catalog}

Once a triggering model has been fitted, the finite-horizon triggering core size can
be computed directly from the observed earthquake history. At an evaluation time
\(t\), each past earthquake \(i\) is assigned its model-implied expected triggering
contribution over \([t,t+T]\). For the temporal Hawkes/ETAS model considered here,
this contribution is
\begin{equation}
 w_{i,T}(t)
 =
 \kappa_i h_T(t-t_i),
\end{equation}
where \(\kappa_i\) is obtained from the fitted productivity relation and the observed
magnitude of earthquake \(i\). The weights are sorted in decreasing order,
\(w_{(1),T}\geq w_{(2),T}\geq\cdots\), and their cumulative sum is compared with
the total weight \(W_T(t)\). The first rank at which the cumulative share reaches
\(p\) gives
\begin{equation}
 N_{p,T}(t)
 =
 \min\left\{
 k:
 \frac{\sum_{j=1}^{k}w_{(j),T}(t)}
 {W_T(t)}
 \geq p
 \right\}.
\end{equation}
Thus, evaluating \(N_{p,T}(t)\) requires neither the reconstruction of parent--offspring
genealogies nor the simulation of future earthquakes. It requires only the observed
catalog and a fitted model that assigns a nonnegative future triggering contribution
to each past earthquake.

Because \(N_{p,T}(t)\) depends on fitted parameters, measured magnitudes, catalog
completeness, and the represented prehistory, its uncertainty should be propagated
rather than reporting only a plug-in estimate. This can be done by recomputing the
weights and their ranking across parameter-bootstrap samples, posterior parameter
draws, magnitude perturbations, and simulated reconstructions of missing events.
Pre-event and post-event values should be distinguished explicitly because the
insertion of the earthquake occurring at \(t\) can change both the total weight and
the triggering core.

For a space--time model and a target region \(\mathcal R\), the event-specific
weight is obtained by integrating the triggering kernel over both the forecast
interval and the region,
\begin{equation}
 w_{i,T,\mathcal R}(t)
 =
 \int_t^{t+T}\int_{\mathcal R}
 g_i(\boldsymbol{x},s\mid\mathcal H_t)\,
 d\boldsymbol{x}\,ds,
\end{equation}
where \(g_i\) is the contribution attributed by the fitted model to past earthquake
\(i\). The same ranking and cumulative-share construction then defines
\(N_{p,T,\mathcal R}(t)\).

\subsection{Complementary effective population sizes}

We complement the finite-horizon triggering core size curve \(p\mapsto\NpT\) with two scalar effective population
sizes that summarize the complete distribution of triggering shares according to standard
statistical measures. These quantities vary continuously with the weights, have a
direct interpretation as equivalent numbers of equally weighted contributors, and respond
differently to dominant and diffuse components of the triggering history.

Let \(s_i=w_{i,T}/\WT\) denote the share of the total finite-horizon triggering potential
carried by earthquake \(i\), so that \(\sum_i s_i=1\). We first define the
inverse-participation-ratio effective population size,
\begin{equation}
 N_{\mathrm{IPR}}
 =\frac{1}{\sum_i s_i^2}.
 \label{eq:ipr}
\end{equation}
This quantity is the order-2 Hill effective number, equivalently the reciprocal of the
quadratic concentration or Simpson index \cite{Simpson1949,Hill1973,Jost2006}. It gives
the number of equally weighted earthquakes that would produce the same value of
\(\sum_i s_i^2\) as the observed distribution of triggering weights. Because the shares
enter quadratically, \(N_{\mathrm{IPR}}\) is particularly sensitive to the largest
contributors.

We also define the entropy-based effective population size,
\begin{equation}
 N_{\mathrm{H}}
 =\exp\left(-\sum_i s_i\log s_i\right).
 \label{eq:hill-shannon}
\end{equation}
This quantity is the order-1 Hill effective number, obtained by exponentiating the
Shannon entropy of the triggering shares \cite{Hill1973,Jost2006}. It gives the number
of equally weighted earthquakes having the same Shannon entropy as the observed
distribution. Compared with \(N_{\mathrm{IPR}}\), \(N_{\mathrm{H}}\) assigns relatively
greater importance to the broader population of small and intermediate contributors.

Both quantities equal \(K\) when the triggering potential is distributed equally among
\(K\) earthquakes, and both approach one when a single earthquake carries nearly all
the triggering weight. They therefore provide continuous complements to \(\NpT\):
their joint use helps distinguish dominance by a small leading set from a more broadly
distributed triggering history.

The three measures obey useful pathwise comparisons. If \(k=N_{p,T}\), the leading
\(k\) shares sum to at least \(p\). Cauchy--Schwarz therefore gives
\begin{equation}
 p^2\leq k\sum_i s_i^2,
 \qquad\text{hence}\qquad
 N_{p,T}\geq p^2N_{\mathrm{IPR}}.
 \label{eq:np_ipr_bound}
\end{equation}
Moreover, monotonicity of Hill numbers with their order implies
\begin{equation}
 N_{\mathrm{IPR}}\leq N_{\mathrm H}.
 \label{eq:hill_order}
\end{equation}
There is no fixed ordering between \(N_{p,T}\) and \(N_{\mathrm H}\), because the former
depends on the chosen retained fraction \(p\), whereas the latter summarizes the complete
share distribution. Equations~\eqref{eq:np_ipr_bound}--\eqref{eq:hill_order} provide exact
checks for numerical and empirical implementations.

These effective population sizes are well defined for an infinite stationary history
whenever the corresponding sums converge. By contrast, the count
\(N_{1,T}\) is meaningful only after specifying a finite catalog or a finite
history cutoff. For an infinite stationary history and a triggering kernel that is
strictly positive at every finite lag, every past earthquake has a positive
finite-horizon weight. Consequently, no finite subset captures exactly all the
triggering potential, and \(N_{1,T}=\infty\) for every \(\theta>0\). This should not
be confused with the divergence of the total infinite-horizon residual triggering
potential for an Omori kernel with \(0<\theta\leq1\).

\section{Continuum and Marked-Poisson Theory}

\subsection{Constant-fertility benchmark}

With \(\kappa\equiv n\), weights are monotone in age for both exponential and power law memory kernels, so the largest weights are
the most recent events. In a finite catalog, let us write the ordered ages as
\(0<a_{[1]}<\cdots<a_{[M]}\). The finite-horizon triggering core size is then exactly
\begin{equation}
 N_{p,T}=\min\left\{k:
 \sum_{j=1}^{k}\hT(a_{[j]})\ge p\sum_{j=1}^{M}\hT(a_{[j]})\right\}.
 \label{eq:finite_catalog_constant}
\end{equation}
The fertility cancels from the shares but still affects the catalog through its branching ratio.
Replacing the random catalog by its stationary rate gives a continuum age cutoff \(A_{p,T}\).
For an exponential memory kernel (\ref{eq:exponential}),
\begin{equation}
 A_{p,T}=-c\log(1-p),
 \qquad
 N_{p,T}^{\mathrm{mf}}=\lambar c\log\frac{1}{1-p},
 \label{eq:exp_constant}
\end{equation}
which is independent of \(T\).

For an Omori memory, let us define \(I(x)=\int_0^x\Phi(u)du\), so that
\begin{equation}
 I(x)=
 \begin{cases}
 \displaystyle \frac{c}{1-\theta}
 \left[\left(1+\frac{x}{c}\right)^{1-\theta}-1\right],&\theta\ne1,\\[8pt]
 \displaystyle c\log\left(1+\frac{x}{c}\right),&\theta=1.
 \end{cases}
 \label{eq:omori_I}
\end{equation}
Let \(A_{p,T}\) denote the age cutoff such that earthquakes with ages
\(a\leq A_{p,T}\) account for a fraction \(p\) of the total finite-horizon
triggering potential. The complementary fraction \(1-p\), contributed by
earthquakes older than \(A_{p,T}\), then satisfies
\begin{equation}
 1-p
 =\frac{I(A_{p,T}+T)-I(A_{p,T})}{I(T)},
 \qquad
 N_{p,T}^{\mathrm{mf}}=\lambar A_{p,T}.
 \label{eq:omori_continuum}
\end{equation}
At fixed finite \(T\) and \(p\uparrow1\),
\begin{equation}
 N_{p,T}^{\mathrm{mf}}
 \sim \lambar\left[\frac{Tc^\theta}{I(T)(1-p)}\right]^{1/\theta}.
 \label{eq:finiteT_asymptotic}
\end{equation}
The exponent is \(1/\theta\). It must not be confused with the infinite-horizon,
\(\theta>1\) exponent \(1/(\theta-1)\), derived together with the constant- and random-fertility
benchmarks in \ref{app:infinite_forecast}.

The forecast horizon continuously connects three concentration questions. At fixed positive age,
\begin{equation}
\begin{gathered}
 \frac{w_{i,T}(t)}{T}\longrightarrow \kappa_i\phi(a_i)\quad(T\downarrow0),
 \\ \qquad
 w_{i,T}(t)\longrightarrow \kappa_i\Phi(a_i)\quad(T\to\infty).
 \label{eq:horizon_hierarchy}
\end{gathered}
\end{equation}
The first limit ranks contributions to the instantaneous endogenous rate. The second ranks total
remaining direct fertility and is normalizable over an infinite stationary past only when
\(\int_0^\infty\Phi(a)da<\infty\). Finite \(T\) is therefore the primary object, while the two
limits are useful benchmarks.

Writing \(N_p^{\mathrm{inst}}(t)\) for the quantile formed from
\(w_i^{\mathrm{inst}}=\kappa_i\phi(a_i)\), and \(N_p^{\mathrm{res}}(t)\) for the
infinite-horizon quantile formed from \(w_i^{\mathrm{res}}=\kappa_i\Phi(a_i)\) when
that normalization exists, their contrast defines
\begin{equation}
 \Delta N_p(t)=N_p^{\mathrm{res}}(t)-N_p^{\mathrm{inst}}(t).
 \label{eq:instantaneous_residual_difference}
\end{equation}
For exponential memory the two normalized share vectors are identical and
\(\Delta N_p=0\). For Omori memory, \(\Phi(a)\) decays one power of age more slowly
than \(\phi(a)\), so the residual statistic gives relatively more weight to remote
earthquakes. No universal sign of \(\Delta N_p\) is asserted for every finite realization,
because fertility heterogeneity and discrete rank changes can dominate locally. The finite-\(T\)
surface avoids the normalization problem of \(N_p^{\mathrm{res}}\) when
\(0<\theta\leq1\).

\subsection{ETAS fertility and deterministic threshold representation}

With a Gutenberg--Richter magnitude density \cite{GutenbergRichter1944} and exponential productivity \cite{Utsu1970a,Utsu1970b},
\begin{equation}
\begin{gathered}
 \kappa(m)=\kappa_0e^{\alpha(m-m_0)},\\ \qquad
 f_M(m)=\beta e^{-\beta(m-m_0)},\qquad
 \gamma=\frac{\beta}{\alpha},
\end{gathered}
\end{equation}
the fertility distribution is Pareto:
\begin{equation}
 \Prob(\kappa>x)=\left(\frac{\kappa_0}{x}\right)^\gamma,
 \quad x\ge\kappa_0,
 \qquad
 n=\frac{\gamma\kappa_0}{\gamma-1},
 \label{eq:pareto}
\end{equation}
requiring \(\gamma>1\) for finite mean fertility.
The truncated fertility first moment reads
\begin{equation}
 \E[\kappa\bm{1}_{\{\kappa>x\}}]=
 \begin{cases}
 n, &0\le x<\kappa_0,\\
 n(\kappa_0/x)^{\gamma-1},&x\ge\kappa_0.
 \end{cases}
 \label{eq:pareto_truncated}
\end{equation}

With constant fertility, the triggering weights decrease monotonically with age, so the
largest contributors can be selected by an age cutoff \(A_{p,T}\). Under ETAS fertility
heterogeneity, this ordering no longer holds: an older but highly fertile earthquake may
carry more finite-horizon weight than a younger earthquake with lower fertility. The
continuum calculation must therefore rank earthquakes according to their complete weight
$w=\kappa h_T(a)$,
which combines age-dependent memory and event-specific fertility.

To implement this ranking, let us replace the discrete stationary history by its continuum
population, with expected number
\(\lambar\,da\) of earthquakes having ages in \([a,a+da]\), and introduce a deterministic
weight threshold \(q\). Earthquakes satisfying \(\kappa h_T(a)>q\) constitute the
continuum analogue of the leading contributors in the ranked discrete history. Their
expected number is
\begin{equation}
 N_T(q)
 =\lambar\int_0^\infty
 \Prob\!\left\{\kappa h_T(a)>q\right\}\,da,
 \label{eq:threshold_number}
\end{equation}
while their expected cumulative finite-horizon triggering weight is
\begin{equation}
 W_T(q)
 =\lambar\int_0^\infty
 h_T(a)\,
 \E\!\left[
 \kappa\bm{1}_{\{\kappa h_T(a)>q\}}
 \right]da.
 \label{eq:threshold_weight}
\end{equation}
In particular, \(W_T(0)\) is the total triggering weight of the continuum population.
We choose \(q_*=q_*(p,T)\) so that the contributors above the threshold carry the
prescribed fraction \(p\) of this total:
\begin{equation}
 \frac{W_T(q_*)}{W_T(0)}=p.
 \label{eq:threshold_selection}
\end{equation}
The corresponding continuum, or mean-field, approximation to the triggering core size
is then
\begin{equation}
 N_{p,T}^{\mathrm{mf}}=N_T(q_*).
 \label{eq:threshold_core}
\end{equation}

This construction thresholds ensemble-averaged weight contributions using the
deterministic value \(q_*\). It should not be confused with the exact ensemble mean
\(\E[\NpT]\). In an individual history, both the total triggering weight and the
threshold required to capture its fraction \(p\) are random and fluctuate jointly.
Consequently, the ratio-of-averages construction above does not, in general, equal the
average of the pathwise triggering core size.

For the exponential memory kernel defined in
equation~\eqref{eq:exponential},
the threshold construction can be evaluated analytically. The forecast-horizon factor
\(1-e^{-\omega T}\) is common to every earthquake, whereas the dependence on age is
entirely contained in \(e^{-\omega a}\). It is therefore convenient to express the
weight threshold in dimensionless form as
\begin{equation}
 x=\frac{q}{\kappa_0(1-e^{-\omega T})}.
\end{equation}
The value \(x=1\) is the weight, in these rescaled units, of a minimum-fertility
earthquake at age zero and separates two selection regimes.

When \(x\geq1\), the threshold exceeds the weight of an age-zero earthquake with the
minimum fertility \(\kappa_0\). An earthquake can then enter the selected triggering
core only through a sufficiently large fertility, with the required fertility increasing
with its age. This regime corresponds to retained triggering fractions
\(0<p\leq p_*\), where
\begin{equation}
 p_*=\frac{1}{\gamma}.
\end{equation}
When \(0<x<1\), every earthquake younger than
\begin{equation}
 a_q=\frac{1}{\omega}\log\frac{1}{x}
\end{equation}
exceeds the threshold even at the minimum fertility, while older earthquakes exceed it
only if their fertility is sufficiently large. This second regime corresponds to
\(p_*<p<1\). Evaluating \(N_T(q)\) and \(W_T(q)\) in these two regimes and eliminating
the threshold \(q\) through \(W_T(q)/W_T(0)=p\) gives the two branches
\begin{equation}
 N_{p,T}^{\mathrm{mf}}=
 \begin{cases}
 \displaystyle
 \frac{\lambar}{\omega\gamma}
 (\gamma p)^{\gamma/(\gamma-1)},
 & 0<p\leq 1/\gamma,\\[7pt]
 \displaystyle
 \frac{\lambar}{\omega}
 \left[
 \log\!\left(\frac{\gamma-1}{\gamma(1-p)}\right)
 +\frac{1}{\gamma}
 \right],
 & 1/\gamma<p<1.
 \end{cases}
 \label{eq:exp_pareto}
\end{equation}
The two expressions coincide at \(p=p_*=1/\gamma\). The first branch describes a
small triggering core selected primarily through unusually large fertilities. In the
second branch, the core includes all sufficiently recent earthquakes together with
older earthquakes selected according to their fertility.

Because \(1-e^{-\omega T}\) multiplies every weight by the same factor, it changes
neither the ordering of the earthquakes nor their normalized shares
\(s_i=w_{i,T}/W_T\). The exponential-memory triggering core size is consequently
independent of the forecast horizon \(T\). This exact factorization is specific to
exponential memory. For an Omori kernel, the finite-horizon weight \(h_T(a)\) does not
separate into a product of a horizon-dependent factor and an age-dependent factor, so
changing \(T\) can alter both the relative weights and their ranking.

For comparison, the infinite-forecast Omori calculation is well defined for an infinite
stationary past only when \(\theta>1\) and is derived in
\ref{app:infinite_forecast}. When \(0<\theta<1\), an infinite forecast instead
requires an explicitly finite historical lookback. The corresponding
finite-lookback limit, presented in \ref{app:lookback}, produces the crossover
\(\theta\gamma=1\). This crossover concerns the joint effects of fertility heterogeneity
and long Omori memory in that distinct limiting construction; it is not a stationarity
condition for the finite-horizon observable studied here.

\subsection{Exact marked-Poisson representation}

As a tractable benchmark, let us replace the Hawkes history by a stationary marked Poisson process with
rate \(\lambar\) and the same fertility law. Mapping age and fertility to
\(w=\kappa\hT(a)\) creates a Poisson point process of weights with tail intensity
\begin{equation}
 \Lambda_T(q)=\lambar\int_0^\infty\Prob\{\kappa\hT(a)>q\}\,da.
 \label{eq:levy_tail}
\end{equation}
Thus, \(\Lambda_T(q)\) is the expected number of triggering weights exceeding
the threshold \(q\):
\begin{equation}
 \Lambda_T(q)
 =\E\!\left[\#\{i:w_i>q\}\right].
\end{equation}
Denoting the intensity density of the weight process by
\(r_T(w)=-\Lambda_T'(w)\), the number of weights exceeding \(q\),
\begin{equation}
 K_q=\#\{i:w_i>q\},
\end{equation}
is Poisson distributed with mean \(\Lambda_T(q)\). For the exact ordered-weight
representation below, we assume that the mapped weight intensity is diffuse, so ties occur
with probability zero. The complementary strictly subthreshold triggering weight,
\begin{equation}
 B_q=\sum_{i:w_i<q}w_i,
 \label{hnthwgb}
\end{equation}
is the total finite-horizon triggering potential carried by earthquakes whose
individual weights lie below \(q\). Its Laplace transform is
\begin{equation}
 \E[e^{-sB_q}]
 =\exp\left[
 -\int_0^q(1-e^{-sw})r_T(w)\,dw
 \right].
 \label{eq:subthreshold_laplace}
\end{equation}

This Laplace transform (\ref{eq:subthreshold_laplace}) determines the distribution of the cumulative weight below any fixed threshold \(q\). 
To connect this subthreshold weight to the triggering core size, let us order the Poisson weights as $Y_1\geq Y_2\geq\cdots$.
and let \(Y_k\) denote the \(k\)th largest weight. Conditional on \(Y_k=y\), define \[ A_{k-1}(y)=\sum_{j=1}^{k-1}Y_j \] 
as the sum of the \(k-1\) weights larger than \(y\). The total weight carried by the \(k\) largest contributors is therefore \[ S_k(y)=y+A_{k-1}(y). \] 
The remaining weight is precisely the strictly subthreshold total $B_q$ (\ref{hnthwgb}), with the threshold set equal to the random 
value \(y\).  Conditional on \(Y_k=y\), the Poisson processes on \((0,y)\) and \((y,\infty)\) 
are independent. Consequently, \(B_y\) is independent of \(A_{k-1}(y)\), and its conditional Laplace transform is 
obtained from equation~\eqref{eq:subthreshold_laplace} by replacing \(q\) with \(y\): 
\begin{equation} 
\E\!\left[e^{-sB_y}\mid Y_k=y\right] =\exp\left[-\int_0^y(1-e^{-sw})r_T(w)\,dw\right]. 
\label{eq:conditional_subthreshold_laplace} 
\end{equation} 

The event \(\NpT\leq k\) occurs when the \(k\) largest weights account for at least the fraction \(p\) of the total weight:
\begin{equation} 
S_k(y)\geq p\bigl[S_k(y)+B_y\bigr]. 
\end{equation} 
Equivalently, 
\begin{equation} 
B_y\leq \frac{1-p}{p}\,S_k(y) =r_p\bigl[y+A_{k-1}(y)\bigr], \qquad r_p=\frac{1-p}{p}. 
\label{eq:core_cdf_condition} 
\end{equation} 

We now derive the distribution of the triggering core size. Its annealed cumulative distribution function is, by definition, 
\begin{equation} 
F_{p,T}^{\mathrm{ann}}(k) =\Prob\{\NpT\leq k\}, 
\label{eq:annealed_cdf_definition} 
\end{equation} 
where the probability is taken over realizations of the marked-Poisson history. Thus, \(F_{p,T}^{\mathrm{ann}}(k)\) 
is the probability that the \(k\) largest triggering weights are sufficient to account for at least the fraction \(p\) 
of the total finite-horizon triggering potential. To evaluate this probability, first fix the \(k\)th largest weight at \(Y_k=y\) 
and the sum of the \(k-1\) larger weights at \(A_{k-1}(y)=a\). The \(k\) largest weights then carry the total $y+a$,
whereas all weights smaller than \(y\) carry the complementary total \(B_y\). As shown in equation~\eqref{eq:core_cdf_condition}, 
these \(k\) largest weights account for at least the fraction \(p\) precisely when 
\[ B_y\leq r_p(y+a), \qquad r_p=\frac{1-p}{p}. \] For fixed \(y\) and \(a\), the only remaining random quantity in this inequality is \(B_y\). 
Therefore, 
\begin{equation} 
\Prob\!\left\{ \NpT\leq k \,\middle|\, Y_k=y,\ A_{k-1}(y)=a \right\} = F_{B_y}\!\left(r_p(y+a)\right), 
\label{eq:conditional_core_cdf} 
\end{equation} 
where \[ F_{B_y}(z)=\Prob\{B_y\leq z\mid Y_k=y\} \] is the CDF of the total weight below \(y\). 
This distribution is determined by the conditional Laplace transform in equation~\eqref{eq:conditional_subthreshold_laplace}. 
The conditioning can now be removed in two stages. First, for a fixed value \(Y_k=y\), 
we average equation~\eqref{eq:conditional_core_cdf} over the random sum \(A_{k-1}(y)\) of the \(k-1\) 
larger weights: 
\begin{equation}
\begin{gathered} 
\Prob\{\NpT\leq k\mid Y_k=y\} \\ = \E\!\left[ F_{B_y}\!\left( r_p\bigl[y+A_{k-1}(y)\bigr] \right) \,\middle|\, Y_k=y \right]. 
\label{eq:core_cdf_given_y} 
\end{gathered}
\end{equation} 
Second, we average this conditional probability over the possible values of the \(k\)th largest weight \(Y_k\). 
This gives 
\begin{equation}
\begin{gathered} 
F_{p,T}^{\mathrm{ann}}(k) \\ = \int_0^\infty f_{Y_k}(y) \E\!\left[ F_{B_y}\!\left( r_p\bigl[y+A_{k-1}(y)\bigr] \right) \,\middle|\, Y_k=y \right]dy. 
\label{eq:exact_poisson_cdf} 
\end{gathered}
\end{equation} 
The density of \(Y_k\) is 
\begin{equation} 
f_{Y_k}(y) = \frac{ e^{-\Lambda_T(y)} \Lambda_T(y)^{k-1}r_T(y) }{(k-1)!}. 
\label{eq:kth_weight_density} 
\end{equation} 
Equation~\eqref{eq:exact_poisson_cdf} is therefore the exact annealed CDF of \(\NpT\) for the 
marked-Poisson benchmark. Its evaluation requires the distribution of the subthreshold total \(B_y\), 
obtained by numerically inverting its Laplace transform, together with an average over the \(k-1\) 
weights larger than \(y\). Alternatively, the same CDF can be estimated directly by simulating the one-dimensional Poisson process of triggering weights.
If a triggering model induces atoms in the weight distribution, an explicit tie-breaking convention
and the corresponding boundary atom must instead be included.

For exponential memory and constant fertility,
\(r_T(w)=\eta/w\) on a bounded interval, with \(\eta=\lambar/\omega\). This scale-invariant
Poisson cloud is connected to generalized Dickman and Poisson--Dirichlet structures
\cite{Arratia1999,Pitman2003}. With Pareto fertilities, the small-weight part remains \(\eta/w\)
and the upper part inherits the Pareto exponent. \ref{app:exponential_cloud} gives the
complete L\'evy density, total-weight transform, and exact single-event condensation probability.

\subsection{Finite-horizon Omori distributional limit}

For Omori memory, \(\hT(a)\sim K_Ta^{-1-\theta}\), where
\(K_T=\theta Tc^\theta\). Let \(\delta=1/(1+\theta)\). Provided
\(\E[\kappa^\delta]<\infty\), the tail intensity measure  (\ref{eq:levy_tail}) reads
\begin{equation}
 \Lambda_T(q)\sim C_Tq^{-\delta}, 
 \qquad
 C_T=\lambar K_T^\delta\E[\kappa^\delta],
 \quad q\downarrow0.
 \label{eq:small_weight}
\end{equation}
This moment always exists for the ETAS Pareto law with \(\gamma>1\), because \(\delta<1<\gamma\).
The cloud of omitted small weights self-averages, while the full \(\WT\) remains random. For the
marked-Poisson benchmark,
\begin{equation}
 (1-p)^{1/\theta}\NpT
 \ \Longrightarrow\
 A_T\WT^{-1/\theta},
 \qquad
 A_T=C_T\left(\frac{C_T}{\theta}\right)^{1/\theta}.
 \label{eq:distributional_limit}
\end{equation}
If the negative moment is finite, then
\(\E[\NpT]\sim A_T(1-p)^{-1/\theta}\E[\WT^{-1/\theta}]\). Jensen's inequality gives
\(\E[\WT^{-1/\theta}]\ge\E[\WT]^{-1/\theta}\), explaining why a deterministic plug-in can
underestimate the annealed mean.

Exponential memory is the marginal small-weight case. It gives
\begin{equation}
 \frac{\NpT}{\log[1/(1-p)]}\xrightarrow{\Prob}\eta=\frac{\lambar}{\omega},
 \qquad p\uparrow1.
 \label{eq:exp_limit}
\end{equation}

\section{Hawkes Characterization and Probability Laws}

For a self-exciting Hawkes/ETAS history, in contrast to the marked-Poisson
benchmark, past earthquakes are organized into branching clusters: immigrants
generate offspring, which may themselves trigger further generations. Consequently,
event times, ages, and finite-horizon weights are statistically dependent, and the
mapped weights \(w_i=\kappa_i h_T(a_i)\) do not form a one-dimensional Poisson point
process. In particular, the independence between weights below and above a threshold
\(q\), used in the exact marked-Poisson representation, no longer holds. This
distinction arises from self-exciting genealogical clustering and is independent of
whether the triggering kernel has exponential or Omori memory.

The Hawkes--Oakes cluster representation nevertheless provides an exact transform of the 
count and cumulative weights above and below any fixed deterministic threshold \(q\). 
We first construct this transform for one cluster and then superpose all immigrant clusters 
to recover the complete stationary history. 

Consider a single immigrant earthquake of age \(u\), 
together with all of its descendants that occurred before the forecast origin. Denote this cluster by \(\mathcal C_u\). 
For each earthquake \(j\in\mathcal C_u\), let \(a_j\) be its age, \(\kappa_j\) its fertility, and $w_j=\kappa_j h_T(a_j)$
its finite-horizon triggering weight. At the fixed threshold \(q\), let us define the cluster-level count above the 
threshold by 
\begin{equation} 
K_{u,q}^{+} = \sum_{j\in\mathcal C_u}\bm{1}_{\{w_j>q\}}, 
\end{equation} 
and the cluster-level cumulative weights above and below the threshold by 
\begin{equation} 
W_{u,q}^{+} = \sum_{j\in\mathcal C_u} w_j\bm{1}_{\{w_j>q\}}, \qquad W_{u,q}^{-} = \sum_{j\in\mathcal C_u} w_j\bm{1}_{\{w_j\leq q\}}. 
\end{equation} 
The joint probability-generating/Laplace functional of these three cluster-level 
quantities is 
\begin{equation} 
G_u(z,s_+,s_-;q) = \E\!\left[ z^{K_{u,q}^{+}} \exp\!\left( -s_+W_{u,q}^{+}-s_-W_{u,q}^{-} \right) \right], 
\label{eq:cluster_functional_definition} 
\end{equation} 
for \(0\leq z\leq1\) and \(s_+,s_-\geq0\). Thus, \(G_u\) describes one cluster whose immigrant root has the prescribed age \(u\). 

To derive an equation for \(G_u\), first consider the contribution of one earthquake 
with weight \(w\). Its count and weight contributions to the transform are represented by 
\begin{equation}
\begin{gathered} 
R_q(w;z,s_+,s_-) \\ = z^{\bm{1}_{\{w>q\}}} \exp\!\left[ -s_+w\bm{1}_{\{w>q\}} -s_-w\bm{1}_{\{w\leq q\}} \right]. 
\label{eq:root_factor} 
\end{gathered}
\end{equation} 
If \(w>q\), this factor contributes one power of \(z\) and assigns \(w\) to the above-threshold total. 
If \(w\leq q\), it contributes no power of \(z\) and assigns \(w\) to the below-threshold total. 

Let us now condition on the fertility \(\kappa\) of the immigrant root. Since the root has age \(u\), 
its weight is \(\kappa h_T(u)\), and its own contribution is \(R_q(\kappa h_T(u);z,s_+,s_-)\). 
Its direct offspring form a Poisson process of delays \(r>0\) with intensity \(\kappa\phi(r)\,dr\). 
Only offspring with \(0<r<u\) occur before the forecast origin. An offspring born after delay \(r\) 
has age \(u-r\) and initiates an independent descendant subtree whose functional is \(G_{u-r}(z,s_+,s_-;q)\). 
Conditional on the root fertility \(\kappa\), let
\(\mathcal O_u(z,s_+,s_-;q\mid\kappa)\) denote the joint transform of all
descendant subtrees initiated by its direct offspring, excluding the contribution
of the root itself. By the probability-generating functional of the Poisson
offspring process,
\begin{equation}
\begin{gathered}
 \mathcal O_u(z,s_+,s_-;q\mid\kappa)
 \\ =
 \exp\left\{
 \kappa\int_0^u
 \phi(r)
 \left[
 G_{u-r}(z,s_+,s_-;q)-1
 \right]dr
 \right\}.
 \label{eq:offspring_functional}
\end{gathered}
\end{equation}
The functional of the complete cluster is obtained by multiplying the root factor
by the transform of its offspring subtrees and then averaging over the root
fertility:
\begin{equation}
\begin{gathered}
 G_u(z,s_+,s_-;q)
 \\ =
 \E_\kappa\!\left[
 R_q\!\left(\kappa h_T(u);z,s_+,s_-\right)
 \mathcal O_u(z,s_+,s_-;q\mid\kappa)
 \right].
 \label{eq:cluster_functional_compact}
\end{gathered}
\end{equation}
Substituting equation~\eqref{eq:offspring_functional} gives the nonlinear Volterra
equation
\begin{widetext}
\begin{align}
 G_u(z,s_+,s_-;q)
 ={}&
 \E_\kappa\Bigg[
 R_q\!\left(\kappa h_T(u);z,s_+,s_-\right)
 \nonumber\\
 &\qquad\times
 \exp\left\{
 \kappa\int_0^u
 \phi(r)
 \left[
 G_{u-r}(z,s_+,s_-;q)-1
 \right]dr
 \right\}
 \Bigg].
 \label{eq:cluster_functional}
\end{align}
\end{widetext}
 This nonlinear Volterra equation determines the joint transform for a single cluster rooted at age \(u\). 

We next pass from one cluster to the complete stationary Hawkes/ETAS history. Let \(K_q^{+}\) 
denote the total number of past earthquakes whose weights exceed \(q\), and let \(W_q^{+}\) and \(W_q^{-}\)
denote the corresponding cumulative weights above and below \(q\), after summing over all 
immigrant clusters: 
\begin{equation} 
K_q^{+} = \sum_{i:t_i<t}\bm{1}_{\{w_i>q\}},
 \end{equation} 
\begin{equation} 
W_q^{+} = \sum_{i:t_i<t}w_i\bm{1}_{\{w_i>q\}}, \qquad W_q^{-} = \sum_{i:t_i<t}w_i\bm{1}_{\{w_i\leq q\}}. 
\end{equation} 
Their joint probability-generating/Laplace functional is defined by 
\begin{equation} 
\mathcal L_q(z,s_+,s_-) = \E\!\left[ z^{K_q^{+}} \exp\!\left( -s_+W_q^{+}-s_-W_q^{-} \right) \right]. 
\label{eq:stationary_functional_definition} 
\end{equation} 
The distinction between the two functionals is therefore explicit: \(G_u\) (\ref{eq:cluster_functional_compact}) describes one cluster 
whose root has age \(u\), whereas \(\mathcal L_q\) (\ref{eq:stationary_functional_definition}) describes the complete stationary history 
obtained by superposing all such clusters. 

Immigrant roots form a Poisson process over past ages \(u>0\) with rate \(\nu\), and the 
clusters generated by distinct immigrants are independent. Applying the Poisson generating-functional 
formula to this process of immigrant clusters gives 
\begin{equation} 
\mathcal L_q(z,s_+,s_-) = \exp\left\{ \nu\int_0^\infty \left[ G_u(z,s_+,s_-;q)-1 \right]du \right\}. 
\label{eq:stationary_functional} 
\end{equation} 
Equations~\eqref{eq:cluster_functional} and \eqref{eq:stationary_functional} therefore characterize, for every fixed \(q\), 
the joint law of the number of weights above \(q\), their cumulative weight, and the cumulative weight below \(q\)
 in a stationary Hawkes/ETAS history. 
 
 The triggering core size \(\NpT\), however, is not defined at a predetermined threshold. For each realization, 
 its threshold is the random weight at which the ranked cumulative contribution first reaches 
 the prescribed fraction \(p\) of the random total weight. Determining the distribution of \(\NpT\) 
 therefore requires the joint evolution of \(K_q^{+}\), \(W_q^{+}\), and \(W_q^{-}\) as the threshold \(q\) varies. 
 Separate inversions of \(\mathcal L_q\) at individual fixed values of \(q\) provide the joint law at each threshold 
 but do not provide this cross-threshold coupling. 
 
 We consequently obtain the distribution of \(\NpT\)
  for stationary Hawkes/ETAS histories by exact Poisson-cluster simulation. The asymptotic 
  amplitude derived for the independent marked-Poisson benchmark is then treated as a prediction 
  to be tested numerically under Hawkes/ETAS clustering, rather than as an already proved result for the self-exciting process.

Three sampling laws must also be distinguished. At a random calendar origin,
\begin{equation}
 F_{p,T}^{\mathrm{ann}}(k)=\Prob\{\NpT(0)\le k\}
 \label{thynbgqq}
\end{equation}
is the annealed stationary CDF of the  triggering core size \(\NpT\). Conditional on a completely known history and fixed parameters,
\(\NpT(t)\) is deterministic; its strict same-time conditional law is a point mass. A nontrivial
single-catalog law is instead the calendar-time occupation CDF of the triggering core size \(\NpT\)
\begin{equation}
 \widehat F_{p,T;L}^{\mathrm{occ}}(k)
 =\frac{1}{L}\int_{t_0}^{t_0+L}\bm{1}_{\{\NpT(t)\le k\}}dt.
 \label{eq:occupation}
\end{equation}
Under stationary ergodicity, equation~\eqref{eq:occupation} converges almost surely to the
annealed CDF (\ref{thynbgqq}). Laws of large numbers and central-limit results for stable Hawkes processes support
this identification under appropriate regularity \cite{Bacry2013,Zhu2013}.

Sampling the process at earthquake occurrence times does not produce the same
distribution as sampling it at uniformly chosen calendar times. The resulting
event-centered distribution is called the \emph{Palm law}: it describes the process
as seen from a typical event, conventionally placed at time zero \cite{DaleyVereJones2008,BaccelliBremaud2003}. We denote expectation
under this law by \(\E^0\), in contrast to the ordinary stationary expectation
\(\E\) at an arbitrary calendar time.

Event-time sampling preferentially selects periods of high seismicity because
earthquakes are more likely to occur when the conditional intensity is large. For a
predictable observable \(g\), evaluated immediately before the event at time zero,
this intensity weighting gives
\begin{equation}
 \E^0[g]
 =
 \frac{\E[\lambda(0^-)g]}{\lambar},
 \label{eq:palm_bias}
\end{equation}
where \(\lambda(0^-)\) is the pre-event conditional intensity and
\(\lambar=\E[\lambda(0)]\) is the stationary mean rate \cite{DaleyVereJones2008,BaccelliBremaud2003}. Thus, the Palm law is an
intensity-weighted version of the calendar-time law: states with larger conditional
intensity are sampled more frequently at event times.

If the observable is evaluated immediately after the event, as in our numerical
experiment, its Palm distribution contains a second effect. In addition to the
pre-event intensity bias, the earthquake at time zero is inserted into the history
and contributes its own finite-horizon weight. The post-event Palm law therefore
differs from the calendar-time law through both preferential sampling of high-intensity
states and the deterministic addition of the newly observed event.

A difference between calendar-time and event-time distributions is consequently
expected even for a stationary and perfectly ergodic process; it should not be
interpreted as ergodicity breaking. Moreover, nearly unstable and heavy-tailed Hawkes
processes may converge very slowly
\cite{JaissonRosenbaum2015,JaissonRosenbaum2016}. 

These considerations separate two sources of discrepancy. First, the
calendar-time occupation distribution estimated from a finite catalog may differ
from the annealed stationary distribution because successive observations are
serially correlated, especially near criticality. Second, the event-time
distribution converges to the Palm law rather than to the calendar-time law and
may therefore remain systematically different even for an arbitrarily long
catalog.
Appendix~\ref{app:occupation_variance} quantifies the first effect. For the
indicator \(X_k(t)=\bm{1}_{\{\NpT(t)\leq k\}}\), it expresses the variance of
the finite-record occupation CDF in terms of the autocovariance of \(X_k\).
The associated integrated correlation time determines the effective number of
independent calendar-time observations and therefore the rate at which the
occupation CDF approaches its annealed limit. The appendix also defines a
blockwise discrepancy statistic that tests whether different portions of a
single catalog converge toward the same annealed distribution. These formulas
provide the uncertainty benchmark needed to distinguish slow finite-record
self-averaging from the persistent calendar-time/event-time difference generated
by Palm sampling.

\section{Simulation Methods}

All Hawkes/ETAS catalogs are generated through the exact Poisson-cluster construction rather
than a discretized intensity. Immigrants form a homogeneous Poisson process. Each event of
fertility \(\kappa_i\) produces a Poisson number of direct children with mean \(\kappa_i\), whose
delays are sampled by inverse transformation from the normalized exponential or Omori kernel.
Children recursively generate later generations. This is the same branching construction used
in ETAS simulation and stochastic reconstruction \cite{HawkesOakes1974,Zhuang2004}.

Fertilities are either constant or Pareto. The finite-horizon ETAS experiments use \(\gamma=1.8\)
and \(\kappa_0=n(\gamma-1)/\gamma\). The finite-lookback crossover study fixes \(\theta=0.5\) and
uses \(\gamma=1.5,2,3\), spanning \(\theta\gamma=0.75,1,1.5\). The main sweeps cover \(n=0.4, 0.7, 0.9\),
\(\theta=0.35, 0.4, 0.6,1, 1.2, 1.5\), multiple \(p\), and \(T=0.3, 3, 30, 300\). Every reported
finite-horizon weight is evaluated from equation~\eqref{eq:weight} and ranked directly.

Table~\ref{tab:simulation_regimes} records the branching ratio for every numerical block.
This distinction is important because the marked-Poisson benchmark has no genealogy: in that
case \(n\) labels the mean fertility only. In the Hawkes experiments, \(n\) is the branching ratio.
Except in the finite-\(T\) constant-fertility amplitude test, the stationary mean rate is held fixed
when \(n\) is varied by adjusting the immigrant rate. The transient branching sweep uses
\(\lambar=0.8\), so its dependence on \(n\) is not confounded by a simultaneous change in mean
event density.

\begin{table*}[!tp]
\caption{Branching and rate settings of the numerical experiments.\label{tab:simulation_regimes}}
\centering
\small
\begin{tabular}{llll}
\toprule
Experiment & Process/fertility & $n$ & $\lambar$ \\
\midrule
Finite-$T$ amplitude & Hawkes/constant & 0.7 & 6.67 \\
Distributional law & marked Poisson/Pareto & $\E[\kappa]=0.7$ & 0.667 \\
Distributional law & Hawkes/Pareto & 0.7 & 0.667 \\
Occupation and Palm & Hawkes/Pareto & 0.4, 0.7, 0.9 & 0.8 \\
Transient condensation & Hawkes/Pareto & 0.4, 0.7, 0.9 & 0.8 \\
Effective populations & Hawkes/Pareto & 0.7 & 0.8 \\
\bottomrule
\end{tabular}
\end{table*}

Slow Omori memory requires explicit prehistory control. For distributional high-\(p\) tests,
histories extend as far as 200,000 time units, and the expected weight of the omitted remote past
is included analytically. A realization is retained for asymptotic comparison only when this
correction is less than one quarter of the target omitted weight \((1-p)\WT\), and when the
selected population is below 20\% of the explicitly represented history.

The distributional experiment uses 300 independent histories per Omori configuration and 1,500
for the exponential marginal test. The occupation experiment uses 600 independent histories and
12 independent length-4096 catalogs per configuration. The stationary total event rate is held at $\lambar = \nu/(1-n)=0.8$ while
\(n\) changes and $\nu$ adapts according to \(\nu=0.8(1-n)\) to keep $\lambar$ constant. Calendar samples lie on a regular grid with spacing 2;
post-event values are evaluated immediately after an event and are never pooled with the calendar
sample. Confidence intervals are computed by resampling independent catalogs, not serially
correlated time points.

For the transient-condensation experiment, we first simulate a stationary history up
to the insertion time and compute its pre-insertion finite-horizon triggering potential,
denoted by \(W_T^{\mathrm{pre}}\). We then insert a new earthquake at the evaluation
time. Because its age is zero, its finite-horizon weight is
\begin{equation}
 w_j=\kappa_j h_T(0)
 =\kappa_j[1-\Phi(T)].
\end{equation}
Immediately after its insertion, the new earthquake alone accounts for at least the
fraction \(p\) of the total triggering potential, and hence produces
\(N_{p,T}=1\), if and only if
\begin{equation}
 w_j
 \geq p\left(w_j+W_T^{\mathrm{pre}}\right).
\end{equation}
Equivalently, its fertility must satisfy
\begin{equation}
 \kappa_j
 \geq
 \kappa_{\mathrm{crit}}
 :=
 \frac{pW_T^{\mathrm{pre}}}
 {(1-p)[1-\Phi(T)]}.
 \label{eq:condensation_threshold}
\end{equation}
The threshold \(\kappa_{\mathrm{crit}}\) is history dependent because it is calculated
from the realized value of \(W_T^{\mathrm{pre}}\). To initialize every realization at
a comparable degree of concentration, we assign the inserted earthquake the fertility
\begin{equation}
 \kappa_j=r\kappa_{\mathrm{crit}},
 \qquad r>1,
 \label{eq:condensation_multiplier}
\end{equation}
where \(r\) controls how far the initial state lies above the boundary
\(N_{p,T}=1\). Thus, \(r=1\) places the new earthquake exactly at the condensation
threshold, whereas larger \(r\) produces stronger initial dominance.

With the ETAS productivity relation
\(\kappa(m)=\kappa_0e^{\alpha(m-m_0)}\), the critical fertility corresponds to the
history-dependent magnitude threshold
\begin{equation}
 m_{\mathrm{crit}}
 =
 m_0+\frac{1}{\alpha}
 \log\left[
 \frac{pW_T^{\mathrm{pre}}}
 {(1-p)\kappa_0[1-\Phi(T)]}
 \right].
 \label{eq:condensation_magnitude}
\end{equation}
The inserted fertility \(r\kappa_{\mathrm{crit}}\) therefore corresponds to
\begin{equation}
 m_j=m_{\mathrm{crit}}+\frac{\log r}{\alpha}.
\end{equation}
We then simulate the complete descendant cluster of the inserted earthquake together
with independent background clusters. A counterfactual simulation with no arrivals
after the insertion time isolates the deterministic redistribution caused by memory
decay from changes produced by newly arriving earthquakes.

\section{Numerical Results}

\subsection{Finite-horizon scaling}

We first test the continuum prediction for constant fertility and Omori memory. 
Equation~\eqref{eq:omori_continuum} determines the age cutoff \(A_{p,T}\) 
for the triggering core: earthquakes younger than \(A_{p,T}\) account for the fraction \(p\) of 
the total finite-horizon triggering potential, while older earthquakes account for the complementary fraction 
\(1-p\). Explicitly, the cutoff satisfies equation~\eqref{eq:omori_continuum}
and the corresponding mean-field triggering core size is \(N_{p,T}^{\mathrm{mf}}=\lambar A_{p,T}\). 
Thus, studying the high-\(p\) behavior of \(N_{p,T}^{\mathrm{mf}}\) amounts to determining how far into the 
past one must extend the triggering core in order to capture an increasingly large fraction of the future 
triggering potential. 

At fixed finite \(T\), equation~\eqref{eq:finiteT_asymptotic} predicts 
$N_{p,T}^{\mathrm{mf}} \sim 1/(1-p)^{1/\theta},  p\uparrow1$.
Consequently, a log--log plot of \(N_{p,T}^{\mathrm{mf}}\) against \(1-p\) must have asymptotic slope \(-1/\theta\), 
or equivalently a plot against \((1-p)^{-1}\) must have slope \(1/\theta\). The physical origin of this scaling is the 
long Omori tail: as \(p\) approaches one, the small omitted fraction \(1-p\) is carried by progressively older 
earthquakes, and the required age cutoff grows as \((1-p)^{-1/\theta}\). Smaller values of \(\theta\) 
therefore produce a much more rapid growth of the triggering core because memory decays more slowly. 

We solved equation~\eqref{eq:omori_continuum} by monotone bisection for \(\theta=0.35,0.6,1,1.5\), 
forecast horizons \(T=0.3,3,30,300\), and \(1-p=10^{-2},\ldots,10^{-8}\). At large \(A_{p,T}\), 
the numerator \(I(A_{p,T}+T)-I(A_{p,T})\) is the difference between two nearly equal quantities. 
Direct subtraction then loses numerical precision, especially for small \(\theta\) and very large \(p\). 
We therefore evaluated this difference using a cancellation-safe representation of the corresponding tail integral. 

The numerical solutions recover the predicted exponents \(1/\theta\) to numerical precision: 
\(2.857143\), \(1.666667\), \(1.000000\), and \(0.6667\) for \(\theta=0.35,0.6,1,\) and \(1.5\), 
respectively. They also converge to the finite-\(T\) prefactor in equation~\eqref{eq:finiteT_asymptotic} 
for every tested horizon. This calculation is a direct numerical verification of the continuum asymptotic formula, 
not a fit to simulated Hawkes catalogs. 

We next test how accurately the continuum prediction describes 
finite stationary Hawkes histories. For each simulated catalog, the empirical triggering core size is 
compared with the continuum value computed at the same \(p\), \(T\), \(\theta\), and stationary rate. 
Because a finite simulation represents only a bounded portion of the past, we exclude cases in which 
the age of the oldest earthquake entering the empirical triggering core exceeds \(20\%\) of the available 
history. This criterion removes the most obvious cases in which the selected core approaches the 
artificial beginning of the simulated catalog. 

The median ratios of the empirical Hawkes core size to 
the continuum prediction are \(0.68\), \(0.96\), \(1.04\), and \(1.19\) for \(\theta=0.35,0.6,1,\) and \(1.5\), 
respectively. Ratios close to one for \(\theta\geq0.6\) show that the continuum calculation captures the 
characteristic scale of the finite-catalog triggering core despite event discreteness and Hawkes clustering. 
The case \(\theta=0.35\) exhibits a substantially smaller ratio because its very slow memory assigns 
appreciable cumulative weight to earthquakes far in the past. Even after applying the \(20\%\) exclusion
 criterion, the finite simulated prehistory omits part of this remote contribution. The empirical triggering 
 core is therefore systematically too small relative to its infinite-history continuum value. 
 
 These results 
 support the predicted finite-horizon exponent \(1/\theta\) and its amplitude for the tested range, while 
showing that direct simulation of the small-\(\theta\) regime requires exceptionally long prehistories. 
 The discrepancy at \(\theta=0.35\) is accordingly interpreted as a finite-history boundary effect rather than as a failure of the continuum scaling law.

\begin{figure}[!tp]
\centering
\includegraphics[width=\columnwidth,height=0.66\textheight,keepaspectratio]{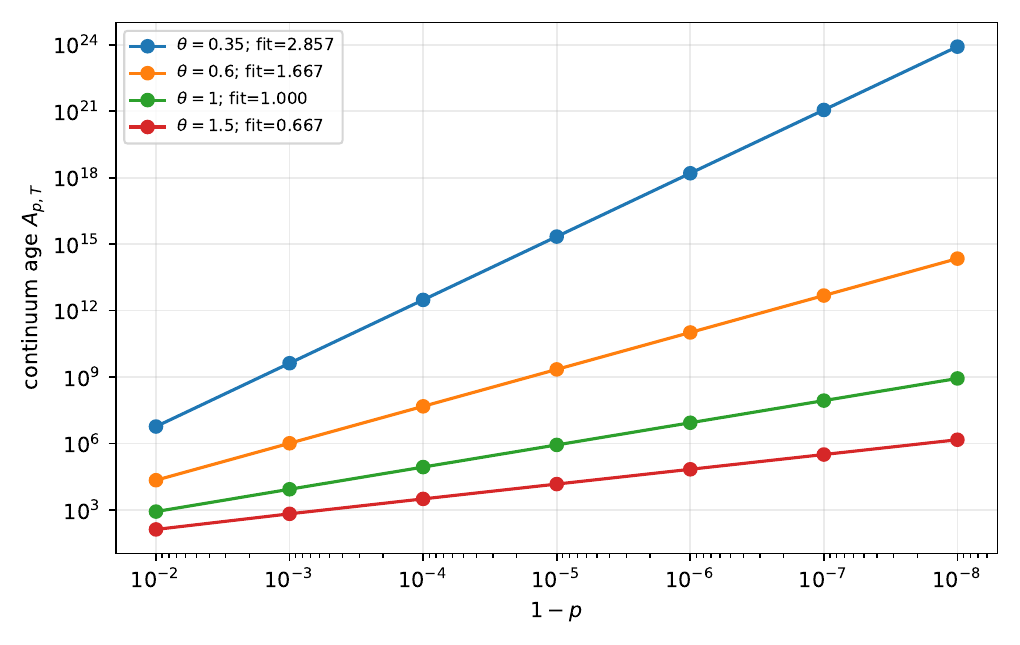}
\caption{Finite-horizon constant-fertility Omori asymptotics. Numerical solutions of
equation~\eqref{eq:omori_continuum} recover the predicted slopes \(1/\theta\) across horizons.
The plotted curves use cancellation-safe tail integrals; direct subtraction is unstable for small
\(\theta\), large ages, and very high \(p\). This deterministic continuum calculation does not
involve a branching ratio.}
\label{fig:finiteT_theory}
\end{figure}

\begin{figure*}[!tp]
\centering
\includegraphics[width=\textwidth,height=0.66\textheight,keepaspectratio]{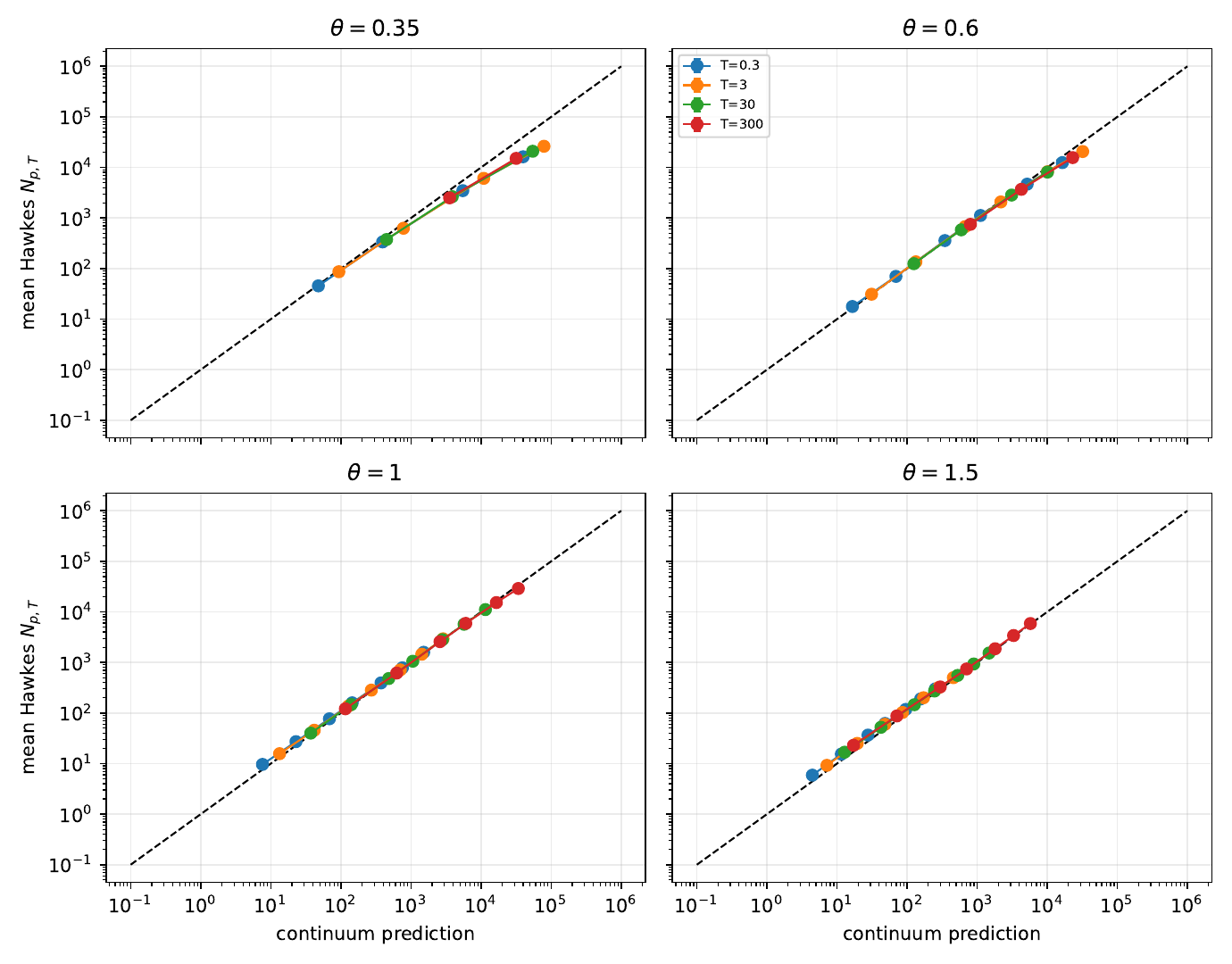}
\caption{Finite-horizon Hawkes verification of the constant-fertility continuum prediction.
Markers summarize 60 independent exact cluster catalogs at \(n=0.7\) and
\(\lambar=6.67\); bars are 95\% Monte Carlo confidence intervals for the mean. The slowest
Omori kernel retains visible history truncation, whereas the remaining exponents approach the
continuum amplitude.}
\label{fig:finiteT_hawkes}
\end{figure*}

Figure~\ref{fig:finiteT_theory} isolates the continuum statement and shows that the
high-\(p\) slopes are independent of branching because no random genealogy enters that
calculation. Figure~\ref{fig:finiteT_hawkes} then tests the amplitude with exact
constant-fertility Hawkes catalogs at \(n=0.7\) and stationary rate
\(\lambar=6.67\). Its main message is twofold: the continuum prediction sets the correct
scale over most of the tested range, while the slowest kernel visibly exposes the finite
left boundary even with a represented prehistory of \(10^5\) time units.

\subsection{Comparison of hard and effective population sizes}

\begin{figure*}[!tp]
\centering
\includegraphics[width=\textwidth,height=0.66\textheight,keepaspectratio]{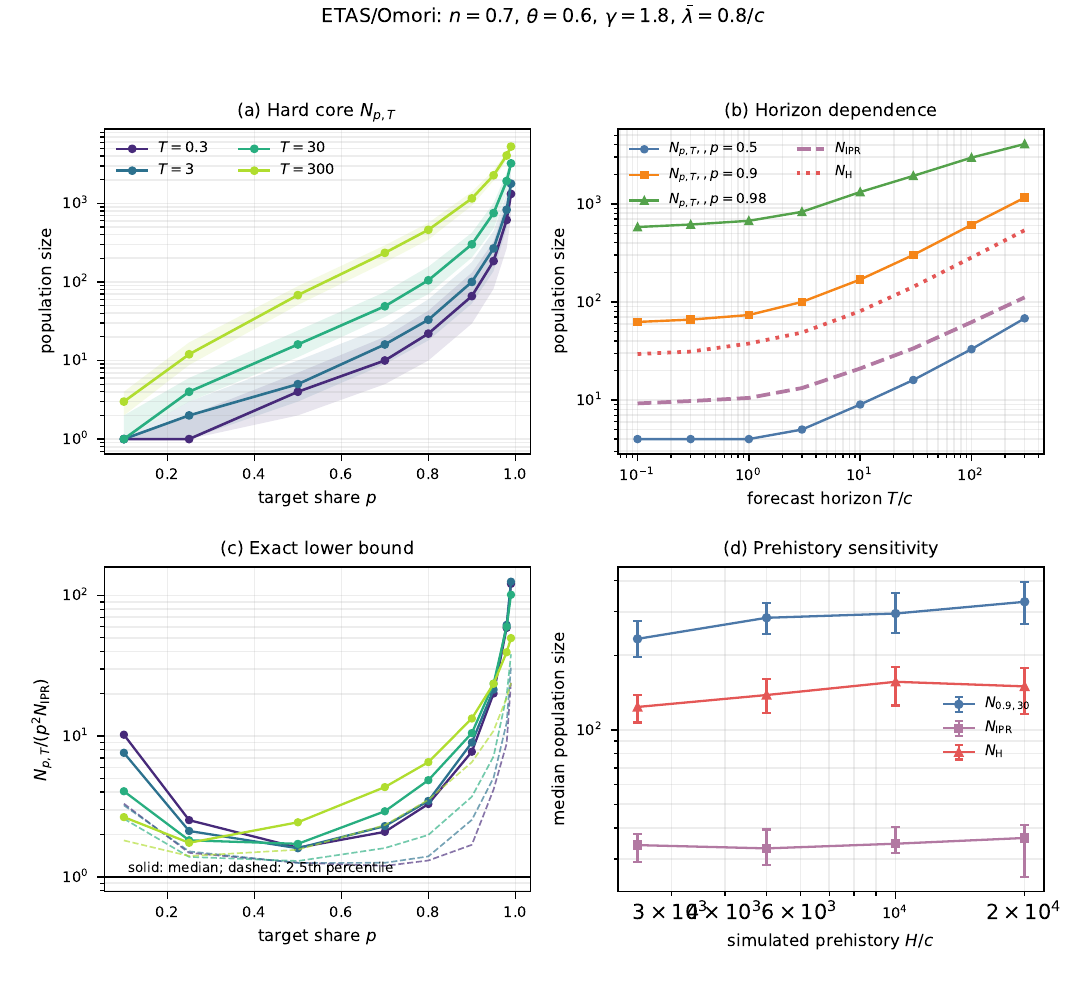}
\caption{Comparison of hard and smooth finite-horizon triggering populations for
ETAS/Omori histories with \(n=0.7\), \(\theta=0.6\), \(\gamma=1.8\), and
\(\lambar=0.8/c\). (a) Median \(N_{p,T}\) as a function of \(p\); shading gives
interquartile realization ranges. (b) Horizon dependence of three hard-core levels together
with \(N_{\mathrm{IPR}}\) and \(N_{\mathrm H}\). (c) Pathwise bound
\(N_{p,T}/(p^2N_{\mathrm{IPR}})\geq1\); solid and dashed curves are medians and 2.5th
percentiles. (d) Prehistory sensitivity at \(p=.9,T=30c\); bars are 95\% bootstrap
confidence intervals for medians from 80 independent catalogs. Panels (a)--(c) use 160
independent histories of length \(10^4c\).}
\label{fig:effective_comparison}
\end{figure*}

We next compare \(N_{p,T}\), \(N_{\mathrm{IPR}}\), and \(N_{\mathrm H}\) on the same
finite-horizon weight vectors. The experiment uses 160 independent ETAS/Omori histories with
\((n,\theta,\gamma,\lambar)=(0.7,0.6,1.8,0.8)\), a represented prehistory
\(H=10^4c\), and common random catalogs across all \(p\) and \(T\). Thus differences between
the measures are not caused by independent Monte Carlo samples. Figure~\ref{fig:effective_comparison}
shows the resulting quantile curves, horizon dependence, exact bound, and prehistory control.

At \(p=0.9\), the median hard core grows from \(66\) at \(T=0.3c\) to \(100\),
\(301\), and \(1155.5\) at \(T=3c,30c,\) and \(300c\), respectively. Over the same
horizons, the median \(N_{\mathrm{IPR}}\) values are \(9.76,13.25,33.59,\) and
\(111.18\), whereas the median \(N_{\mathrm H}\) values are \(31.17,48.73,142.55,\)
and \(537.54\). The ordering \(N_{\mathrm{IPR}}\leq N_{\mathrm H}\) therefore holds
with a substantial gap: the quadratic measure is controlled mainly by the largest weights,
while the entropy measure retains much more information about the diffuse small-weight cloud.

The hard core answers a different, explicitly decision-dependent question. At \(T=30c\),
its median is \(16\) for \(p=0.5\), \(301\) for \(p=0.9\), and \(1925.5\) for
\(p=0.98\). Consequently, \(N_{p,T}\) can lie below or above either smooth effective
population depending on the target share. Across every realization, \(p\), and \(T\), the
implementation satisfies the exact lower bound \(N_{p,T}\geq p^2N_{\mathrm{IPR}}\) from
equation~\eqref{eq:np_ipr_bound}; panel~(c) displays both the median ratio and its 2.5th
percentile without a violation.

Long Omori memory leaves measurable finite-history drift. In an independent 80-catalog
sensitivity experiment at \(p=0.9,T=30c\), increasing \(H/c\) from \(2500\) to \(20000\)
raises the median \(N_{p,T}\) from \(233\) to \(329\). The corresponding medians change from
\(34.14\) to \(36.49\) for \(N_{\mathrm{IPR}}\), and from \(123.57\) to \(149.68\) for
\(N_{\mathrm H}\). The hard high-share core is therefore the most sensitive of the three to
remote omitted history. The entropy effective number is also history sensitive, whereas the
IPR is comparatively stable because very small old weights contribute little after squaring.

\subsection{Distributional predictions}

We now test the distributional scaling law in equation~\eqref{eq:distributional_limit}. 
For the marked-Poisson benchmark, this equation predicts that, as \(p\uparrow1\), 
the rescaled triggering core size \((1-p)^{1/\theta}N_{p,T}\) converges in distribution 
to the random amplitude \(A_TW_T^{-1/\theta}\). The factor \((1-p)^{-1/\theta}\) describes the 
universal growth of the core at high \(p\), \(A_T\) contains the deterministic dependence on
 the forecast horizon and model parameters, and \(W_T^{-1/\theta}\) retains the realization-to-realization 
 fluctuations of the total triggering weight. We test both the realization-level agreement between the two 
 sides and the agreement between their full distributions. 
 
 For the marked-Poisson ETAS/Pareto benchmark at \(T=30\), 
 the median of the realization-level ratio $\frac{(1-p)^{1/\theta}N_{p,T}} {A_TW_T^{-1/\theta}}$ is \(0.995\) at 
 \((\theta, p)=(0.4, 0.9)\), \(0.988\) at \((0.6, 0.95)\), \(0.997\) at \((0.6, 0.98)\), and \(0.985\) at \((1, 0.99)\). 
 Values close to one show that the random amplitude \(A_TW_T^{-1/\theta}\) predicts not only the scaling 
 exponent but also the realization-specific magnitude of the triggering core. Two-sample Kolmogorov--Smirnov 
 distances between the empirical distribution of the rescaled core size and the predicted random-amplitude 
 distribution range from \(0.027\) to \(0.037\), indicating close agreement between the complete distributions. 
 
 The convergence is asymptotic in \(p\), rather than an exact equality at arbitrary quantiles. For example, 
 at \(\theta=1\), the median realization-level ratio rises from \(0.835\) at \(p=0.9\) to \(0.993\) at \(p=0.995\). 
 The discrepancy at the lower value of \(p\) therefore decreases as the triggering core is required to capture 
 a larger fraction of the total weight, as predicted by the high-quantile limit. 
 
 We also test whether \(A_T\) 
 correctly removes the dependence on the forecast horizon. At \(\theta=0.6\) and \(p=0.95\), the median 
 ratios are \(0.998\), \(0.988\), and \(0.986\) for \(T=3\), \(30\), and \(300\), respectively. The nearly 
 unchanged ratios show that the normalization collapses horizons spanning two orders of magnitude 
 onto the same asymptotic amplitude. The more extreme case \(p=0.98\), \(T=300\) is not reported because 
 fewer than half of the simulated histories satisfy the predeclared remote-history criterion. In that regime, 
 the combination of a long forecast horizon and a high retained fraction makes the result too sensitive to 
 the finite beginning of the simulated history. 
 
 The marked-Poisson result is proved using the independent 
 Poisson structure of the weights. We next examine whether the same leading scaling remains informative 
 for self-exciting Hawkes/ETAS histories, where genealogical clustering creates dependence between weights. 
 For \(n=0.7\), representative median ratios are \(0.956\) for \((\theta, p)=(0.4, 0.9)\), \(0.981\) for \((0.6, 0.98)\), 
 and \(0.944\) for \((1, 0.995)\).  Their 
 proximity to one shows that, over the parameter range examined, the
marked-Poisson random-amplitude scaling also provides an accurate leading
approximation for the simulated Hawkes/ETAS histories. Because its analytical
derivation relies on independent Poisson weights, however, the extension to
self-exciting histories is supported here numerically rather than established
theoretically.
 
 The simulations also distinguish the exact annealed mean 
 from the deterministic mean-field approximation. For constant fertility and \(p\) between \(0.9\) and \(0.95\), 
 the exact marked-Poisson annealed mean exceeds the mean-field value by approximately \(6\%\). This 
 difference arises because the realization-dependent threshold and total weight fluctuate jointly: averaging 
 the nonlinear pathwise core size is not equivalent to replacing the random quantities by their means. 
 
 Finally, exponential memory belongs to the marginal small-weight regime, for which equation~\eqref{eq:exp_limit} 
 predicts the slower logarithmic growth $N_{p,T}\sim\eta\log\frac{1}{1-p},  \eta=\frac{\lambar}{\omega}$. 
 For \(\eta=10\) and \(p=0.99999\), the empirical values of \(\E[N_{p,T}]/\log[1/(1-p)]\) are \(10.055\) for constant 
 fertility and \(9.916\) for Pareto fertility. Both values lie within approximately \(1\%\) of the predicted limit \(10\). 
 Their agreement also confirms that, for exponential memory, the leading high-\(p\) coefficient depends on 
 \(\lambar/\omega\) but not on the fertility-mark distribution.

\begin{figure*}[!tp]
\centering
\includegraphics[width=\textwidth,height=0.66\textheight,keepaspectratio]{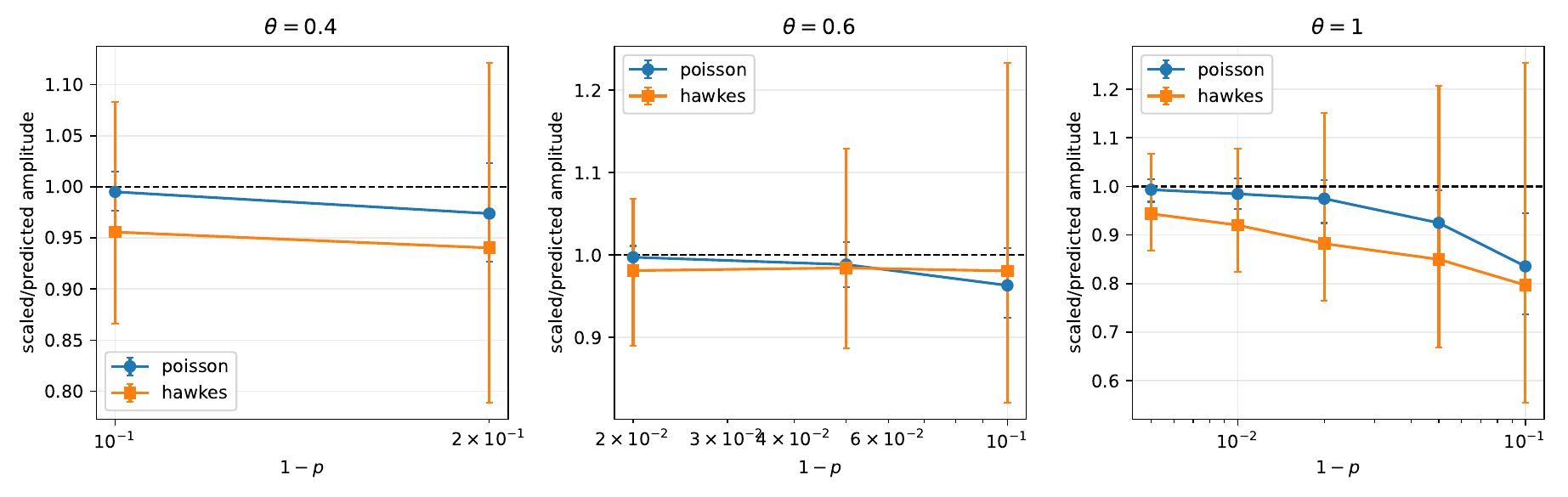}
\caption{Random-amplitude test for finite-horizon Omori weights. Points give the median ratio
between \((1-p)^{1/\theta}N_{p,T}\) and \(A_TW_T^{-1/\theta}\); bars are interquartile ranges.
The marked-Poisson benchmark has mean fertility \(0.7\). Exact Hawkes catalogs use branching
ratio \(n=0.7\) and \(\lambar=0.667\). The Poisson result converges to one; Hawkes catalogs
show a similar but broader and slower collapse, which is numerical evidence rather than a proved
extension. Each Omori setting uses 300 independent histories.}
\label{fig:distributional_amplitude}
\end{figure*}

\begin{figure*}[!tp]
\centering
\includegraphics[width=0.92\textwidth,height=0.66\textheight,keepaspectratio]{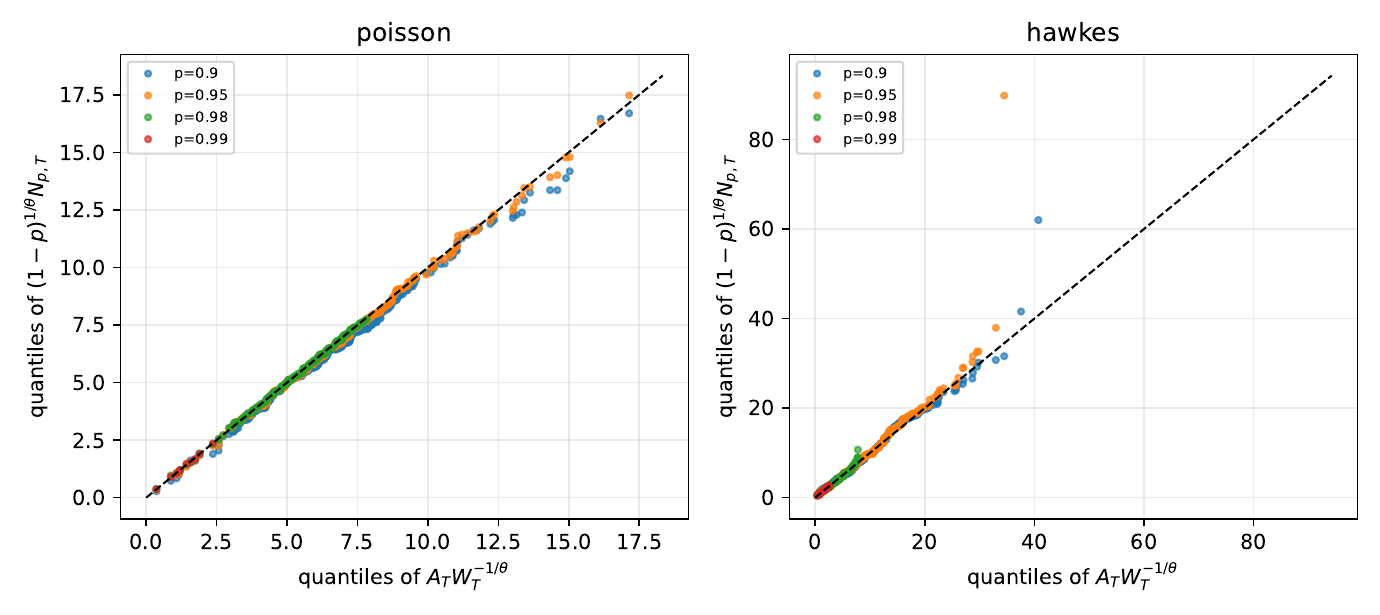}
\caption{Quantile--quantile comparison of the two sides of
equation~\eqref{eq:distributional_limit} for \(\theta=.6,T=30\). The Poisson benchmark follows
the diagonal closely. Hawkes/ETAS histories have \(n=0.7\) and \(\lambar=0.667\): clustering
broadens the finite-sample law but preserves the leading collapse over the tested range.}
\label{fig:distributional_qq}
\end{figure*}

Figure~\ref{fig:distributional_amplitude} summarizes the realization-level amplitude
test and shows where the Hawkes law is broader than the Poisson benchmark. Both use mean
fertility \(0.7\); for the Hawkes histories this is the branching ratio \(n=0.7\), with
\(\lambar=0.667\). Figure~\ref{fig:distributional_qq} then compares the full quantiles,
rather than only a median ratio. Its close Poisson diagonal validates the derived limiting CDF,
while the visible Hawkes broadening quantifies the genealogical correction that is not contained
in the independent-weight proof.

\subsection{Annealed law, occupation law, and Palm bias}

We next examine whether the distribution obtained from a single stationary catalog approaches 
the annealed distribution constructed from independent histories. For a calendar block of duration \(L\), 
we compare its occupation CDF with the annealed CDF. For a calendar block \([t_0,t_0+L]\), the occupation CDF
$\widehat F_{p,T;L}^{\mathrm{occ}}(k)  =
 \frac{1}{L}
 \int_{t_0}^{t_0+L}
 \bm{1}_{\{\NpT(t)\leq k\}}\,dt
$
is the fraction of the block during which the triggering core size does not
exceed \(k\). It is therefore the empirical CDF obtained by sampling a single
catalog uniformly in calendar time.
 If the process self-averages, the discrepancy 
between these two distributions should decrease as progressively longer portions of the catalog are observed. 

This behavior is found for every subcritical configuration tested. Between \(L=64\) and \(L=4096\), 
the mean squared difference between the block and annealed CDFs decreases from \(0.0269\) to 
\(4.44\times10^{-4}\) for Omori memory with \(n=0.4\), from \(0.0354\) to \(2.46\times10^{-3}\) for \(n=0.7\), 
and from \(0.0434\) to \(2.96\times10^{-3}\) for \(n=0.9\). For exponential memory with \(n=0.7\), 
it decreases from \(0.0107\) to \(4.13\times10^{-4}\). Thus, longer calendar blocks progressively recover 
the annealed law, and we find no evidence over the tested durations for a persistent realization-specific 
occupation distribution. Convergence nevertheless becomes markedly slower as the branching ratio approaches one. 

The origin of this slow convergence is the strong temporal dependence of the concentration observable. 
To quantify temporal dependence, let
$ k_{1/2}^{\mathrm{ann}}
 =
 \min\left\{
 k:F_{p,T}^{\mathrm{ann}}(k)\geq\frac{1}{2}
 \right\}
$
denote the median of the annealed distribution, and define the binary process
$ X(t)
 =
 \bm{1}_{\{\NpT(t)\leq k_{1/2}^{\mathrm{ann}}\}}.
$
The integrated correlation time of \(X(t)\) measures how long the process
typically retains information about whether its triggering core size lies below
or above the annealed median.

The integrated correlation times of this median-threshold indicator are
\(16.4\,[13.6, 19.9]\), \(33.8\,[25.9, 41.3]\), and \(70.9\,[52.2, 94.3]\) for Omori memory with \(n=0.4, 0.7,\) 
and \(0.9\), respectively. By comparison, exponential memory with \(n=0.7\) gives the much shorter 
correlation time \(5.25\,[4.94, 5.61]\). Brackets denote \(95\%\) catalog-bootstrap confidence intervals. 
Long Omori memory and near-critical branching therefore cause successive calendar observations 
to carry strongly overlapping information. 

This effect substantially reduces the effective sample size. 
A catalog of duration \(4096\) with Omori memory and \(n=0.9\) contains approximately 
$\frac{4096}{2\times70.9}\simeq29$ effectively independent observations of the indicator, 
even though the regular sampling grid contains more than \(2000\) evaluation times. The large 
number of recorded values therefore overstates the amount of independent information available 
for estimating the occupation law. This provides a concrete explanation for the slow blockwise 
convergence observed near criticality. 

Event-time sampling answers a different question and does 
not converge to the calendar-time law. It produces the Palm distribution, which describes the process 
as seen from a typical earthquake. The Kolmogorov--Smirnov distances between the post-event 
Palm and calendar-time distributions are \(0.285\), \(0.340\), and \(0.431\) for Omori memory with 
\(n=0.4, 0.7,\) and \(0.9\), respectively, and \(0.251\) for exponential memory with \(n=0.7\). 
The increasing separation across the Omori cases shows that event-time bias becomes more 
pronounced as clustering strengthens. 

For Omori memory, post-event sampling shifts the distribution 
toward smaller triggering core sizes, corresponding to stronger concentration. At \(n=0.7\), 
the pooled calendar-time median is \(416\), whereas the post-event median is \(223\). For exponential memory, 
the shift has the opposite direction: the median increases from \(2\) under calendar-time sampling to \(4\) 
immediately after events. Palm sampling is therefore not a universal transformation toward stronger concentration. 
Its net effect combines two mechanisms: event times preferentially sample histories with high pre-event intensity, 
and post-event evaluation inserts the new earthquake and its finite-horizon weight. Their balance depends on 
the memory kernel and the organization of the pre-existing triggering weights. 

The decreasing block-to-annealed 
discrepancy and the persistent Palm/calendar difference thus describe two distinct phenomena. Calendar-time 
occupation laws self-average slowly because of temporal correlation, whereas event-time observations converge 
to a genuinely different, event-centered sampling law. The latter difference is an expected Palm bias and should not be interpreted as ergodicity breaking.

\begin{figure*}[!tp]
\centering
\includegraphics[width=\textwidth,height=0.66\textheight,keepaspectratio]{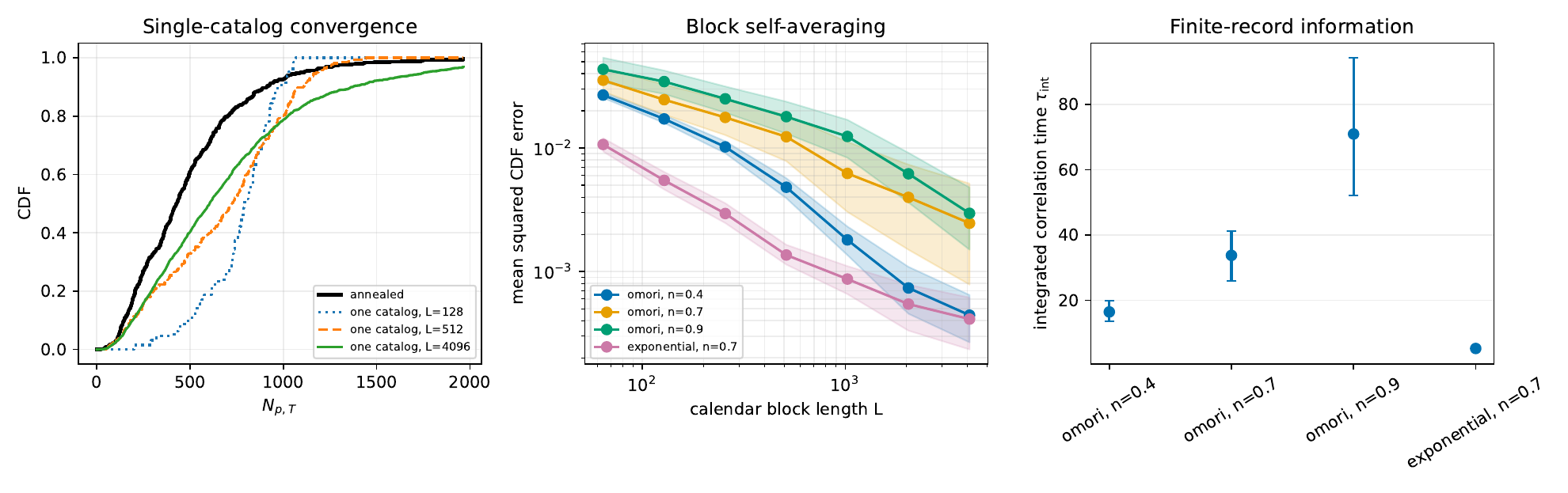}
\caption{Annealed and single-catalog laws for \(p=.9,T=30\) at fixed
\(\lambar=0.8\). (a) A single Omori \(n=.7\)
catalog approaches the independent-history CDF as its duration increases. (b) Mean squared block
CDF error decreases with block length; shaded regions are 95\% catalog-bootstrap intervals.
(c) Integrated correlation time of an indicator thresholded at the annealed median for
\(n=.4,.7,.9\). Near-critical Omori memory sharply reduces the effective amount of independent
information.}
\label{fig:occupation}
\end{figure*}

\begin{figure*}[!tp]
\centering
\includegraphics[width=0.92\textwidth,height=0.66\textheight,keepaspectratio]{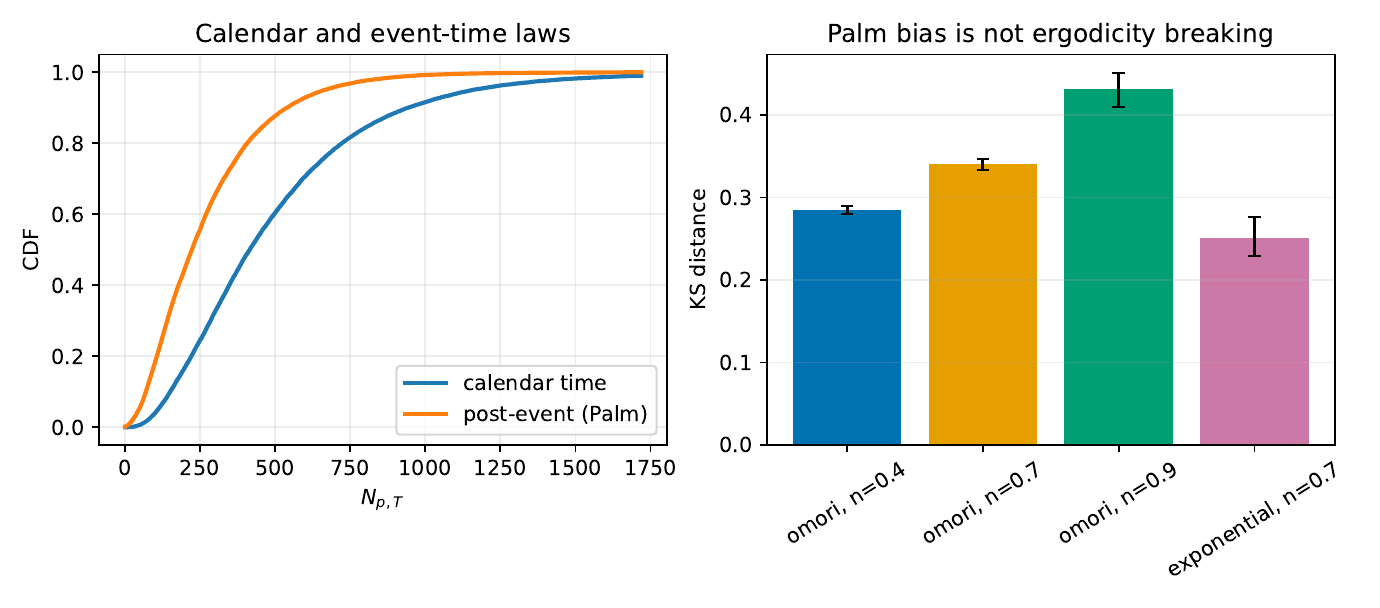}
\caption{Calendar-time and post-event concentration laws. Omori cases shift toward smaller
\(N_{p,T}\), whereas the exponential case shifts toward larger values under the post-event
convention. All cases have \(\lambar=0.8\); Omori branching ratios are \(n=.4,.7,.9\), and
the exponential comparison has \(n=.7\). Catalog-bootstrap intervals show that the
Palm/calendar difference persists as the calendar occupation law self-averages; it is therefore
not an ergodicity-breaking signature.}
\label{fig:palm}
\end{figure*}

Figure~\ref{fig:occupation} connects these conclusions directly to branching strength: at fixed
\(\lambar=0.8\), increasing \(n\) from \(0.4\) to \(0.9\) slows blockwise convergence and
increases the integrated correlation time without changing the mean event rate. Figure~\ref{fig:palm}
shows the complementary sampling result. The calendar-time law self-averages, but its limit is
not the post-event Palm law; the sign of the shift also changes between Omori and exponential
memory.

\subsection{Transient condensation and rank dynamics}

We now examine how a strongly concentrated triggering state relaxes over time. In each stationary realization, 
an earthquake is inserted with fertility \(\kappa_j=r\kappa_{\mathrm{crit}}\), where the history-dependent threshold 
\(\kappa_{\mathrm{crit}}\) is defined by equation~\eqref{eq:condensation_threshold}. For \(r>1\), the inserted 
earthquake initially carries at least the fraction \(p\) of the total finite-horizon triggering weight, so that \(N_{p,T}=1\) 
immediately after insertion. Scaling the inserted fertility to the triggering potential of the realized prehistory gives 
every catalog the same initial concentration criterion and avoids conditioning on an arbitrary and noisy magnitude bin. 

We define the first-exit time from this condensed state as 
\begin{equation} 
\tau_{\mathrm{exit}} = \inf\left\{ \tau>0:N_{p,T}(t_0+\tau)>1 \right\}, 
\label{eq:first_exit_time} 
\end{equation} 
where \(t_0\) is the insertion time. The exit occurs when the inserted earthquake no longer accounts by 
itself for the fraction \(p\) of the total triggering weight. It can result either from the deterministic evolution 
of the existing weights or from the arrival of new earthquakes and their descendants. 

For the baseline parameters 
\((p, T, n, r)=(0.9, 30, 0.7, 1.5)\), the median first-exit time is \(0.359\) for exponential memory. For Omori memory, 
the corresponding medians are \(0.163\), \(0.118\), and \(0.0776\) for \(\theta=0.35, 0.6,\) and \(1.2\), respectively. 
All cases use the common-rate design \(\lambar=0.8\). The distinctive exponential behavior follows from 
the factorization of the exponential finite-horizon weights: between earthquake arrivals, all existing weights are 
multiplied by the same decay factor, so their normalized shares remain unchanged. The system can therefore 
leave the initial \(N_{p,T}=1\) state only when a new earthquake arrives. Its exit time is consequently strongly 
affected by the random waiting time and productivity of subsequent events. For Omori memory, relative 
weights also evolve continuously between arrivals, providing an additional mechanism for leaving the condensed state. 

To separate these two mechanisms, we construct a no-new-arrivals counterfactual in which the pre-existing 
earthquakes continue to age after \(t_0\), but no new earthquakes or descendants are allowed to occur. Under 
exponential memory, this counterfactual produces neither an exit from \(N_{p,T}=1\) nor a change in the weight 
ranking through the maximum tested time \(300\). This follows exactly from the common exponential decay factor. 
In contrast, every tested Omori realization exits the condensed state before time \(300\), even without new arrivals. 
The median counterfactual exit times are \(1.49\), \(1.09\), and \(0.52\) as \(\theta\) increases from \(0.35\) to \(0.6\) 
and \(1.2\). Faster Omori decay therefore accelerates the redistribution of triggering responsibility away from the 
initially dominant earthquake. 

The distance above the initial condensation threshold controls how long this dominance 
persists. Increasing the fertility multiplier \(r\) from \(1.05\) to \(1.5\) and \(3\) increases the median no-arrival exit time 
from \(0.086\) to \(1.09\) and \(5.31\), respectively. A mainshock placed only slightly above \(\kappa_{\mathrm{crit}}\) 
loses its dominant status rapidly, whereas a substantially more fertile earthquake remains the sole member of the 
triggering core for a longer period. 

The complete Hawkes/ETAS process adds the competing effect of newly triggered 
earthquakes. At the fixed stationary rate \(\lambar=0.8\), raising the branching ratio from
\(n=0.4\) to \(n=0.9\) shortens the median first-exit time from \(0.163\) to \(0.0863\).
Because the immigrant rate is reduced as \(n\) rises, this comparison isolates the reorganization
of genealogy and temporal dependence rather than a trivial increase in the number of events.
More of the fixed-rate activity belongs to descendant cascades at large \(n\), and the newly
arriving descendant weights dilute the share carried by the inserted earthquake more rapidly.

The contrast between exponential and Omori memory also determines 
the dynamics of the weight ranking. For exponential memory, $h_T(a)=(1-e^{-\omega T})e^{-\omega a},$ so that, 
between arrivals, every pairwise ratio \(w_i/w_j\) remains constant. Ranks can change only when a new earthquake 
enters the history. The finite-horizon Omori kernel does not factor into a common time-dependent multiplier. 
Different earthquake ages therefore imply different relative decay rates, allowing an older but more fertile earthquake 
to overtake a younger earthquake as both age. Rank crossings are thus an intrinsic feature of deterministic Omori 
memory evolution, not solely a consequence of new arrivals. Because the arrival-driven transitions in the present 
experiment are located on a discrete evaluation grid, their counts are used descriptively rather than for event-exact 
transition-rate inference. 

For the infinite-horizon Omori residual weights, the crossing can be written explicitly. If
\(A_i=\kappa_i^{1/\theta}\) and \(A_i\neq A_j\), solving \(w_i(t)=w_j(t)\) gives
\begin{equation}
 t_{ij}^{\mathrm{cross}}
 =\frac{A_it_j-A_jt_i}{A_i-A_j}-c,
 \label{eq:omori_crossing_time}
\end{equation}
provided this time is later than both event times. For finite \(T\), the corresponding equation
uses \(h_T\) and is solved numerically; the existence of crossings remains, but the simple
closed form does not.

The continuous effective population sizes provide a complementary view of this relaxation. 
Immediately after insertion of the context-scaled Omori earthquake, \(N_{\mathrm{IPR}}\) and \(N_{\mathrm H}\) 
both lie near their single-dominant-earthquake limit of one. They subsequently increase as triggering weight spreads 
from the inserted earthquake to the pre-existing history and to its descendants. The increase of \(N_{\mathrm H}\) 
is larger because the entropy-based effective number is more sensitive than \(N_{\mathrm{IPR}}\) to the diffuse population 
of small triggering shares. The hard triggering core size \(N_{p,T}\) identifies when one earthquake ceases to carry 
the prescribed fraction \(p\), while the two continuous measures reveal how the remaining triggering responsibility 
is progressively redistributed across the broader earthquake population.

\begin{figure}[!tp]
\centering
\includegraphics[width=\columnwidth,height=0.66\textheight,keepaspectratio]{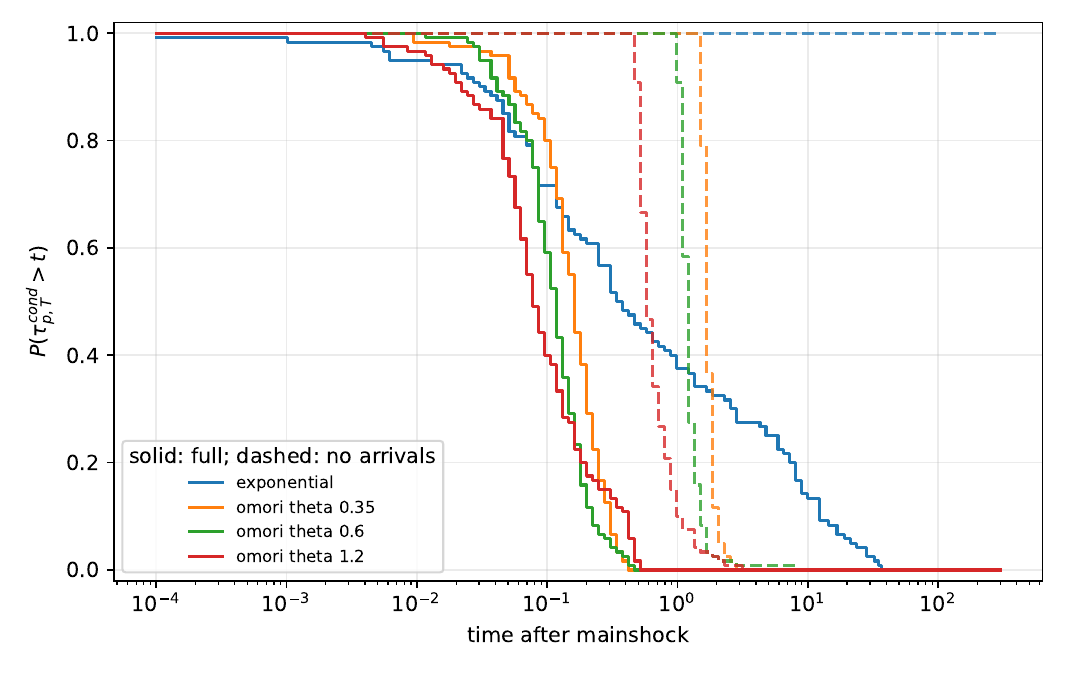}
\caption{Survival of the initial \(N_{p,T}=1\) state following a context-scaled mainshock.
Curves use 120 independent catalogs at \(n=.7,\lambar=.8,p=.9,T=30,r=1.5\).
The no-arrival comparison isolates continuous Omori decondensation from jumps caused by new
events; pointwise uncertainty is summarized by the corresponding realization fractions.}
\label{fig:condensation}
\end{figure}

\begin{figure*}[!tp]
\centering
\includegraphics[width=\textwidth,height=0.66\textheight,keepaspectratio]{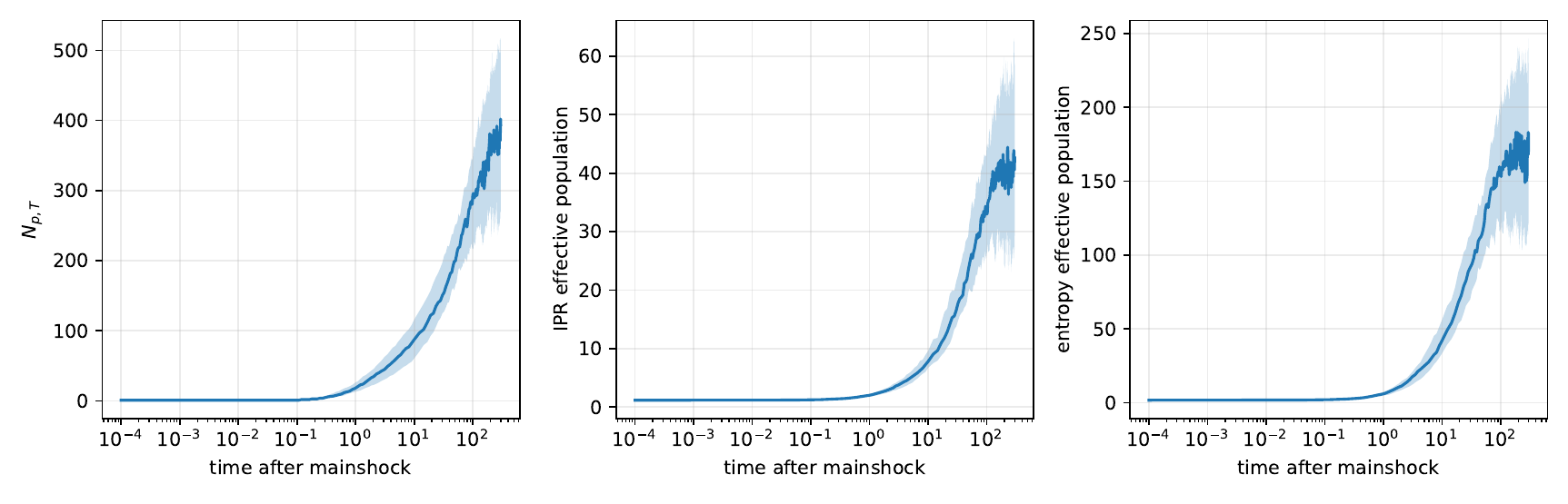}
\caption{Median concentration measures after the baseline Omori mainshock
\((\theta,n,p,T,r)=(.6,.7,.9,30,1.5)\). Shading gives interquartile ranges across 120 independent
catalogs. The hard quantile, inverse-participation effective population, and entropy effective
population all record decondensation, while retaining different sensitivity to small residual
weights.}
\label{fig:effective_condensation}
\end{figure*}

\begin{figure*}[!tp]
\centering
\includegraphics[width=\textwidth,height=0.66\textheight,keepaspectratio]{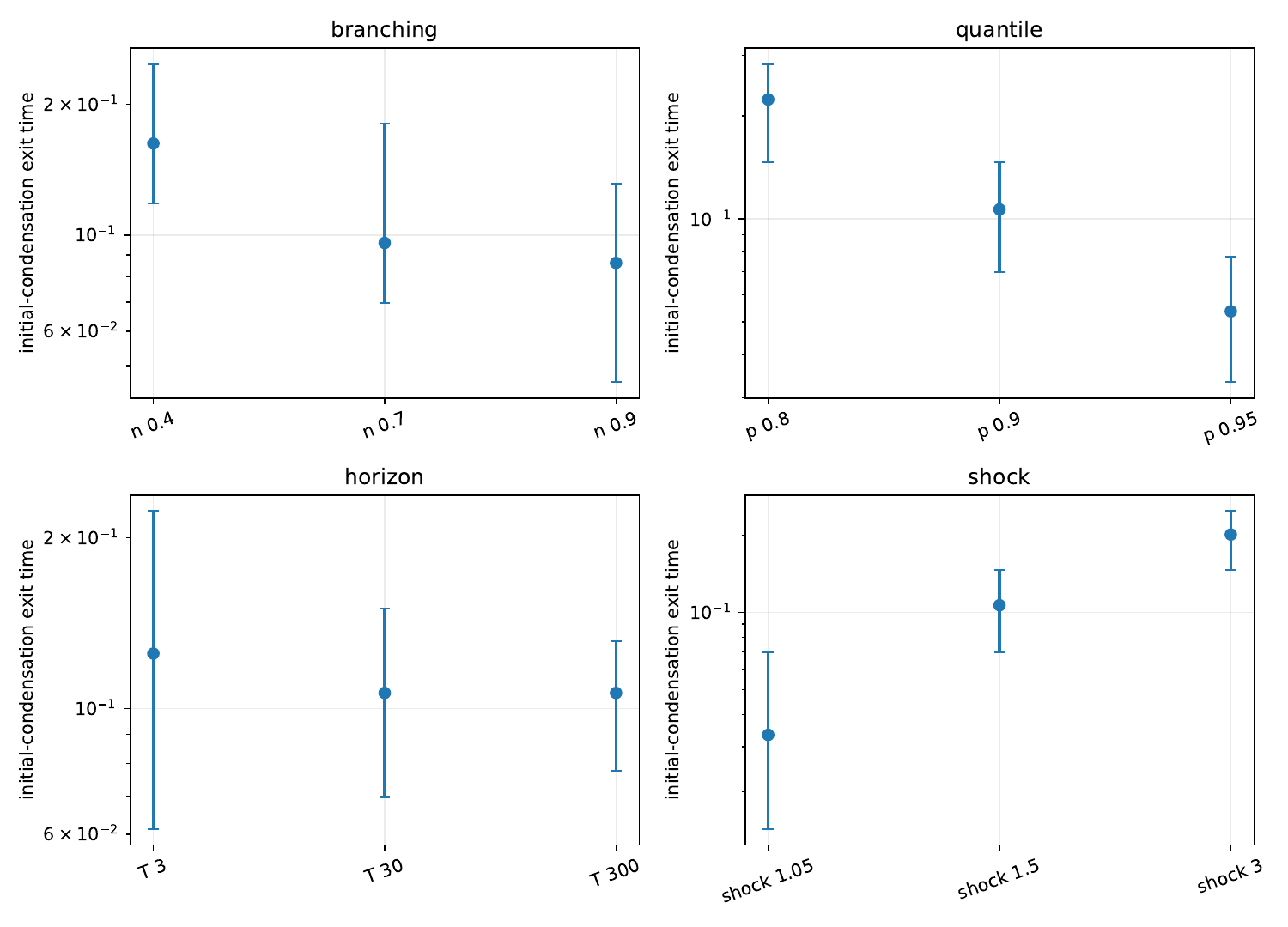}
\caption{One-factor transient-condensation sweeps for the Omori baseline
\((\theta,\gamma,\lambar)=(.6,1.8,.8)\), based on 120 independent catalogs per setting.
Points are median first-exit times and bars span the 25th--75th percentiles. The branching-ratio
sweep varies \(n=.4,.7,.9\) at fixed \(\lambar\); the other panels vary \(p\), \(T\), or the
fertility multiplier \(r\) while holding \(n=.7\).}
\label{fig:condensation_sweeps}
\end{figure*}

Figure~\ref{fig:condensation} separates continuous Omori relaxation from arrival-driven
decondensation. Figure~\ref{fig:effective_condensation} shows that all three population measures
record the same loss of single-event dominance but at different sensitivities. The one-factor
sweeps in Figure~\ref{fig:condensation_sweeps} then connect the exit time to the branching ratio,
target share, horizon, and distance \(r\) above the condensation threshold; all branching cases
use the same \(\lambar=0.8\).

\subsection{Illustrative calculation for the 2016--2017 Central Italy sequence}

To demonstrate that the observable can be evaluated on an actual earthquake history, we use
the open CAT5 machine-learning catalog of the Amatrice--Visso--Norcia sequence
\cite{Tan2021Amatrice,Tan2021AmatriceDataset,Chiaraluce2022Catalogues}. The catalog contains
approximately \(9\times10^5\) detections. We retain the 1298 events with \(M_w\geq3\), the
conservative completeness threshold previously adopted for magnitude-dependence analysis of
this catalog \cite{PetrilloZhuang2023Magnitude}. The calculation uses the full 354.6-day
record and evaluates the weights on a regular calendar grid rather than only at event times.

For this illustrative application, we evaluate the concentration measures using
a preliminary temporal ETAS parameterization of the CAT5 sequence. The calculation
is used only to demonstrate the pathwise evolution of the proposed observables
around the major earthquakes. A full space-time calibration accounting for short-term detection effects, spatial boundaries, and parameter uncertainty is beyond the scope of the present theoretical study.

\begin{figure*}[!tp]
\centering
\includegraphics[width=\textwidth,height=0.66\textheight,keepaspectratio]{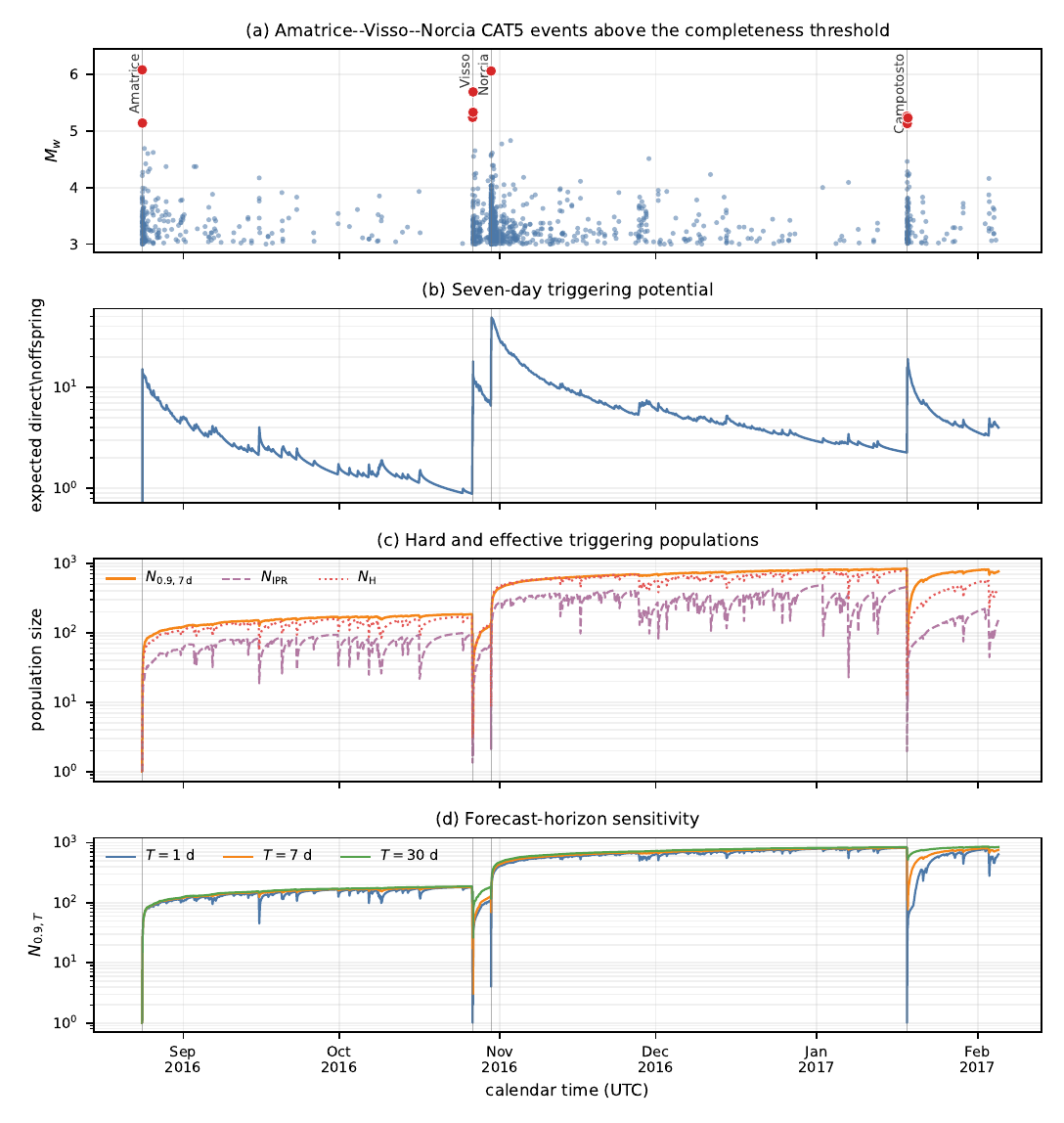}
\caption{Illustrative pathwise triggering concentration for the 2016--2017
Amatrice--Visso--Norcia CAT5 catalog at \(M_w\geq3\). (a) Retained earthquakes,
with major events highlighted. (b) Seven-day finite-horizon triggering potential.
(c) \(N_{0.9,T}\), \(N_{\mathrm{IPR}}\), and \(N_{\mathrm H}\) for \(T=7\) days.
(d) \(N_{0.9,T}\) for \(T=1,7,\) and \(30\) days. The calculation illustrates the pathwise behavior of the observables and is not used for regional parameter
inference or stationary model assessment.}
\label{fig:amatrice_demo}
\end{figure*}

Figure~\ref{fig:amatrice_demo} shows the calculation. Each major earthquake produces a sharp
increase in the seven-day expected number of direct offspring, but the associated concentration
response depends on the existing history. Immediately after the first Amatrice event, the resulting
weight vector gives \(N_{0.9,7\mathrm d}=N_{\mathrm{IPR}}=N_{\mathrm H}=1\). Immediately
after the Visso event the corresponding values are \(4\), \(1.70\), and \(3.53\). After Norcia,
the hard 90\% core is \(69\), while \(N_{\mathrm{IPR}}=2.10\) and \(N_{\mathrm H}=8.58\):
one source dominates the quadratic concentration, yet many small weights are still needed to
capture 90\% of the total. This is precisely the distinction between the hard and effective
populations quantified in Figure~\ref{fig:effective_comparison}. The lower panel further shows
that the inferred core is a forecast-timescale-dependent state variable rather than a permanent
classification of the sequence.

\section{Discussion}

\subsection{New information provided by the finite-horizon triggering core size}

The finite-horizon triggering core size $N_{p,T}(t)$ addresses a question that is
not answered by the ETAS conditional intensity. The intensity $\lambda(t)$ gives the
total instantaneous earthquake rate implied by the observed history, but it does not
reveal how that rate is distributed among past earthquakes. By contrast,
$N_{p,T}(t)$ is the minimum number of past earthquakes whose expected direct
offspring account for the fraction $p$ of the triggering potential over the future
interval $[t,t+T]$. It therefore resolves the internal organization of the
endogenous rate: a small value identifies a compact triggering core dominated by a
few earthquakes, whereas a large value identifies a diffuse state in which triggering
responsibility is spread across a broader history.

This allows one to distinguish two histories that can have the same current
conditional intensity and nevertheless imply different subsequent rate trajectories,
sensitivities, and persistence. If triggering is concentrated in one recent earthquake,
the near-future forecast may depend strongly on its measured magnitude, estimated
productivity, and short-term catalog completeness, and its contribution may decay
rapidly. If the same endogenous rate is supported by many earthquakes of different
ages, no single event controls the forecast, but the combined contribution may persist
for longer and depend more strongly on the accurate representation of the extended
history. The pair
$
 \bigl(\lambda(t),N_{p,T}(t)\bigr)
$
therefore separates the \emph{level} of expected seismic activity from the
\emph{concentration} of its triggering sources. Neither coordinate can in general be
inferred from the other.

The full curve $p\mapsto N_{p,T}(t)$ further describes how the triggering core
expands as increasingly small contributions are included. The complementary effective
population sizes $N_{\mathrm{IPR}}$ and $N_{\mathrm H}$ summarize the same weight
distribution from different perspectives: $N_{\mathrm{IPR}}$ emphasizes the largest
contributors, whereas $N_{\mathrm H}$ is more sensitive to the diffuse population
of smaller shares. Jointly, these observables distinguish a single dominant
earthquake, a compact group of leading earthquakes, and a broad triggering background
that may have similar values of $\lambda(t)$.

The finite forecast horizon is essential both operationally and mathematically. The
weight of earthquake $i$,
$
 w_{i,T}(t)=\kappa_i h_T(a_i),
$
is its conditional expected number of direct offspring in $[t,t+T]$. It can be
computed directly from the observed history and fitted ETAS parameters, without
genealogical reconstruction or simulation of future catalogs. Moreover, the total
finite-horizon weight remains finite for every $\theta>0$ in a stationary
finite-rate process, including the slowly decaying Omori regime
$0<\theta\leq1$, for which the corresponding infinite-forecast residual triggering
potential is not integrable over an infinite stationary past.

The analytical and numerical results identify how the triggering core depends on
memory, fertility heterogeneity, and the forecast horizon. For finite-horizon Omori
memory, the high-$p$ core size grows as $(1-p)^{-1/\theta}$. Slowly decaying
memory therefore requires increasingly deep histories to capture the last fraction of
the triggering potential. For exponential memory, the growth is only logarithmic,
and the common exponential decay factor preserves normalized shares and ranks between
earthquake arrivals. Omori weights do not share such a common factor: relative shares
can evolve continuously, dominant events can lose their position without new arrivals,
and ranks can cross as earthquakes age. The transient-condensation experiments make
this difference visible dynamically by following the relaxation from an initially
single-event triggering core.

The finite-horizon triggering core size also provides a quantitative language for describing the time-dependent roles conventionally denoted by foreshock, mainshock, and aftershock. ETAS does not treat these as intrinsically different event types: their labels depend on their positions within a realized cascade. Similarly, \(N_{p,T}(t)\) does not assign a permanent class to an earthquake, but measures how triggering responsibility is organized at a specified time and forecast horizon. A new earthquake produces a mainshock-dominated triggering state when it alone accounts for at least the fraction \(p\) of the future direct triggering potential, so that \(N_{p,T}=1\). The subsequent increase of \(N_{p,T}\) quantifies the decondensation of this state as responsibility spreads from the initial earthquake to its aftershock cascade. Conversely, the evolution of \(N_{p,T}\) before a later large earthquake can characterize whether the preceding seismicity was organized around a compact triggering core or distributed across many sources. Such preceding events become foreshocks only retrospectively, however, and a concentrated pre-event triggering core is not by itself a prediction of a future mainshock. Establishing whether the statistic contains prospective foreshock information beyond the fitted ETAS intensity requires a dedicated empirical forecast test.

For the independent marked-Poisson benchmark, the high-$p$ theory predicts not only
the scaling exponent but also the realization-dependent amplitude
$A_TW_T^{-1/\theta}$. The numerical experiments reproduce both components across
the tested values of $p$ and $T$. The same leading normalization remains accurate
for the simulated Hawkes/ETAS histories over the parameter range examined. Because
the proof uses the independent Poisson structure of the mapped weights, however, its
extension to self-exciting histories is established here numerically rather than
theoretically. A formal derivation would need to control the coupled evolution of the
thresholded counts and weights as $q$ varies, rather than only their fixed-threshold
marginal transforms in
equations~\eqref{eq:cluster_functional}--\eqref{eq:stationary_functional}.

The dependence on the branching ratio must be interpreted together with the rate convention.
At fixed immigrant rate \(\nu\), stationarity gives \(\lambar=\nu/(1-n)\); consequently,
continuum core counts proportional to \(\lambar\) grow as \((1-n)^{-1}\) when
\(n\uparrow1\). This density effect is absent in our branching-ratio sweeps, which hold
\(\lambar\) fixed by setting \(\nu=\lambar(1-n)\). Any remaining dependence on \(n\)
then reflects changes in genealogy, temporal clustering, fluctuations, and mixing rather than a
trivial increase in the number of earthquakes per unit time.

The distribution of $N_{p,T}(t)$ also provides a reference against which an observed
state can be placed. Defining
$
 U_{p,T}(t)
 =
 F_{p,T}^{\mathrm{ann}}\!\left(N_{p,T}(t)\right)
$
locates the current triggering core size within the annealed distribution implied by a
fitted model. This separates an unusually concentrated or diffuse triggering
configuration from a merely high or low seismicity rate. The occupation and Palm
experiments show, however, that this comparison depends on the sampling convention.
Calendar-time occupation laws approach the annealed law under subcritical
self-averaging, although convergence can be very slow near criticality, whereas
event-time sampling converges to a distinct Palm law because earthquakes
preferentially occur during high-intensity states and post-event evaluation inserts a
new triggering weight.

A convenient real-time anomaly score is
\begin{equation}
 A_{p,T}(t)=-\log_{10}\!\left[\max\{U_{p,T}(t),u_{\min}\}\right],
 \label{eq:anomaly_score}
\end{equation}
with a small numerical floor \(u_{\min}\). This is a state diagnostic, not a forecast
probability. The genuinely conditional future law is instead
\begin{equation}
 F_{p,T;s}^{\mathrm{pred}}(k\mid\mathcal H_t)
 =\Prob\!\left[N_{p,T}(t+s)\leq k\mid\mathcal H_t\right],
 \qquad s>0,
 \label{eq:predictive_concentration_law}
\end{equation}
which includes both future arrivals and deterministic aging of the known weights. It can be
estimated by conditional ETAS simulation from the observed history and calibrated with a
discrete probability-integral transform. Equations~\eqref{eq:anomaly_score} and
\eqref{eq:predictive_concentration_law} separate the diagnosis of the present concentration
state from prediction of its future evolution.

\subsection{Implications for seismicity analysis, forecasting, and limitations}

The principal practical contribution of $N_{p,T}(t)$ is to expose the source
structure underlying an ETAS forecast. Alongside the total rate, it identifies whether
the forecast is supported by one dominant earthquake, a compact recent sequence, or a
large population distributed across the catalog. This information can improve the
interpretation of temporal variations in seismicity. A rise in $\lambda(t)$ caused
by one dominant source is different from an equally large rise produced by many
moderate sources, just as a decline in $N_{p,T}(t)$ signals condensation of
triggering responsibility rather than necessarily an increase in the total rate.
Tracking both quantities can therefore reveal changes of seismic organization that
are invisible in the conditional intensity alone.

The triggering core also provides a direct guide to forecast sensitivity. When
$N_{p,T}(t)$ is small, uncertainty in a few magnitudes, productivity parameters, or
recent detection probabilities can control a large fraction of the endogenous
forecast. These leading contributions can be identified and subjected to targeted
perturbation or scenario analysis. When $N_{p,T}(t)$ is large, sensitivity is less
localized, but errors in the long historical record or in the memory kernel can
accumulate across many small contributions. The metric thus indicates whether forecast
uncertainty is concentrated in a few influential earthquakes or distributed over the
catalog. This distinction can guide uncertainty propagation, robustness checks, and
the allocation of attention to recent-event measurements versus long-history
completeness.

A possible route to forecast improvement is therefore diagnostic rather than an
automatic modification of the ETAS rate. Under a correctly specified ETAS model with
known parameters and a complete history, $N_{p,T}(t)$ is derived from the same
components that determine the conditional intensity and does not by itself add new
events or information. Its value is that it stratifies forecasts according to their
internal concentration. Forecast errors, residuals, and calibration can be tested
separately in concentrated and diffuse regimes, or as functions of the annealed
percentile $U_{p,T}(t)$. Systematic differences would reveal aspects of the data
that are not captured by the fitted ETAS null, such as productivity variability,
magnitude dependence, catalog incompleteness, or an inadequate memory kernel. Whether
conditioning on these regimes yields measurable prospective forecast gains is an
empirical question that should be assessed by out-of-sample likelihood, calibration,
and information-gain tests.

Real-catalog applications should evaluate $N_{p,T}(t)$ on a regular calendar grid
and report pre-event and post-event values separately. Rolling CDF discrepancies
should be calibrated using complete ETAS simulations or block resampling rather than
an independent-sample Kolmogorov law, because successive observations may be strongly
correlated
\cite{Ogata1988,Schoenberg2003,Clements2011,Schorlemmer2007}. Near criticality, a
densely sampled catalog can contain only a small number of effectively independent
concentration states. Event-time distributions require a separate Palm reference and
should not be compared directly with calendar-time percentiles.

Catalog incompleteness following large earthquakes is a particularly important
confounder. Missing small earthquakes remove many weak contributions and can make the
observed triggering history appear artificially concentrated. At the same time,
biased estimates of productivity, memory, and branching alter both the observed
weights and their annealed reference distribution
\cite{SornetteWerner2005b,Seif2017}. Spatial boundaries, magnitude-dependent
detection, uncertainty in event magnitudes, and truncation of the remote past must
therefore be propagated into $N_{p,T}(t)$, rather than treated only as uncertainty
in $\lambda(t)$. These effects are central to any regional application of the
metric.

Several conceptual limitations must be mentioned. First, critical whole-line constructions at $n=1$
require separate existence and ergodicity analyses; the present simulations and analysis make no
claim about that case (see \cite{SaichevSornette2014,Sornette2026CriticalHawkes}).
The finite-lookback concentration crossover \(\theta\gamma=1\) in
Appendix~\ref{app:lookback} must also be distinguished from the heuristic critical whole-line
boundary \(\theta=\gamma-1\) proposed for \(1<\gamma<2\). Their algebraic intersection is
\begin{equation}
 \gamma=\frac{1+\sqrt5}{2},
 \qquad
 \theta=\frac{\sqrt5-1}{2}.
 \label{eq:critical_line_intersection}
\end{equation}
This golden-ratio pair only marks where two different asymptotic criteria coincide; it is not a
stationarity boundary of the finite-horizon statistic, which is well defined for every finite
\(T\) and \(\theta>0\).
Second, $N_{p,T}(t)$ attributes the conditional expectation of \emph{direct}
offspring in $[t,t+T]$. It is not based on the expected number of all descendants
generated within that interval. The latter depends on future generations and cannot
be assigned to the currently observed earthquakes using
equation~\eqref{eq:weight} alone. Both quantities are meaningful, but they answer
different questions: the present observable resolves responsibility for direct
near-future triggering, whereas an all-generation observable would describe the
future propagation of complete branching cascades.
Finally, under the standard ETAS assumption that child magnitude is independent of
parent magnitude and history, multiplying every weight by the probability that a child
exceeds a target magnitude leaves all normalized shares unchanged. A
magnitude-conditioned $N_{p,T}$ surface is therefore flat under this null.
Parent--offspring magnitude dependence, investigated using declustering and
conditional magnitude models
\cite{NicholsSchoenberg2014,Nandan2019,Nandan2022}, can break this invariance.
Departures from the predicted flat surface could consequently provide a focused
diagnostic of magnitude-dependent triggering, but their interpretation and forecasting
value require a dedicated empirical analysis.

\section{Conclusions}

Finite-horizon triggering concentration supplies information absent from the ETAS conditional
intensity: it measures how responsibility for near-future direct triggering is distributed over
the observed earthquake history. The observable is mathematically well posed for slow Omori
memory, has tractable continuum and Poisson limits, and can be evaluated exactly from a catalog
and fitted parameters.

The numerical study validates the finite-horizon Omori exponent and amplitude, the full
marked-Poisson random-amplitude law, and the exponential logarithmic limit. Exact Hawkes/ETAS
simulations reveal additional cluster broadening while supporting the leading high-quantile
collapse. Long-catalog experiments show self-averaging toward the annealed law, strongly slowed
near criticality, and a distinct event-time Palm law. Transient experiments further demonstrate
that Omori memory produces continuous redistribution and rank crossing, whereas exponential
shares change only at arrivals.

These results establish a controlled theoretical and numerical basis for subsequent regional
catalog diagnostics. The next empirical step is to combine \(\lambda(t)\), \(N_{p,T}(t)\),
effective population sizes, and annealed percentiles while explicitly propagating parameter,
history, and detection uncertainties.

\appendix

\section{Infinite-Forecast Continuum Benchmarks}
\label{app:infinite_forecast}

This appendix collects the \(T\to\infty\) formulas used as benchmarks for the finite-horizon
theory in Sections~2 and 3. Its purpose is to show exactly which closed forms survive when the
entire future reproductive potential is integrated, and why these formulas require stronger
memory integrability than the finite-\(T\) observable studied in the main text. We derive both branches of
the ETAS/Pareto threshold construction so that the crossover in \(p\) and the high-\(p\)
exponents quoted in the main text are explicit rather than assumed.

Let \(T=\infty\), so that \(w_i=\kappa_i\Phi(a_i)\). These benchmarks are defined on an infinite
stationary past when \(\int_0^\infty\Phi(a)da<\infty\). For constant fertility and exponential
memory, a continuum age cutoff gives
\begin{equation}
 N_p^{\mathrm{const,exp}}=\frac{\lambar}{\omega}\log\frac{1}{1-p}.
 \label{eq:app_const_exp}
\end{equation}
For Omori memory with \(\theta>1\),
\begin{equation}
 N_p^{\mathrm{const,Omori}}=\lambar c
 \left[(1-p)^{-1/(\theta-1)}-1\right].
 \label{eq:app_const_omori}
\end{equation}
Thus the infinite-forecast high-\(p\) exponent is \(1/(\theta-1)\), rather than the finite-horizon
exponent \(1/\theta\).

For Pareto ETAS fertility and exponential memory, absorb the common finite-horizon factor into
\(r=q/[\kappa_0(1-e^{-\omega T})]\). When \(r\ge1\), equations
\eqref{eq:threshold_number} and \eqref{eq:pareto_truncated} give
\begin{equation}
 N_T(q)=\frac{\lambar}{\omega\gamma}r^{-\gamma},
 \qquad
 W_T(q)=\frac{\lambar n(1-e^{-\omega T})}{\omega\gamma}
 r^{-(\gamma-1)}.
 \label{eq:app_exp_upper}
\end{equation}
Dividing by \(W_T(0)=\lambar n(1-e^{-\omega T})/\omega\) gives
\(p=\gamma^{-1}r^{-(\gamma-1)}\), which produces the first branch of
equation~\eqref{eq:exp_pareto}. When \(0<r<1\), the threshold crosses the minimum fertility at
\(a_q=\omega^{-1}\log(1/r)\), and
\begin{equation}
 N_T(q)=\frac{\lambar}{\omega}\left(\log\frac{1}{r}+\frac{1}{\gamma}\right),
 \qquad
 \frac{W_T(q)}{W_T(0)}=1-\frac{\gamma-1}{\gamma}r.
 \label{eq:app_exp_lower}
\end{equation}
This yields the second branch and shows directly why all normalized exponential results are
independent of \(T\).

For Pareto ETAS fertility, set \(r=q/\kappa_0\). With Omori memory and \(\theta>1\), the
threshold-above-cutoff branch \(r\ge1\) gives
\begin{align}
 N(q)&=\frac{\lambar c}{\theta\gamma-1}r^{-\gamma},\\
 W(q)&=\frac{\lambar n c}{\theta\gamma-1}r^{-(\gamma-1)}.
 \label{eq:app_omori_upper}
\end{align}
Since \(W(0)=\lambar n c/(\theta-1)\), define
\begin{equation}
 p_*=\frac{\theta-1}{\theta\gamma-1}.
\end{equation}
Solving \(W(q)/W(0)=p\) yields
\begin{equation}
 N_p^{\mathrm{ETAS,Omori}}=
 \frac{\lambar c}{\theta\gamma-1}
 \left(\frac{p}{p_*}\right)^{\gamma/(\gamma-1)},
 \qquad 0<p\le p_*.
 \label{eq:app_omori_lowp}
\end{equation}

For \(0<r<1\), the threshold crosses the minimum fertility at
\(a_q=c(r^{-1/\theta}-1)\). Direct integration of the fully selected younger ages and the
Pareto-selected older ages gives
\begin{align}
 N(q)&=\lambar c\left[
 \frac{\theta\gamma}{\theta\gamma-1}r^{-1/\theta}-1\right],\\
 \frac{W(q)}{W(0)}&=1-Ar^{(\theta-1)/\theta},
 \qquad A=\frac{\theta(\gamma-1)}{\theta\gamma-1}.
 \label{eq:app_omori_lower}
\end{align}
Therefore
\begin{equation}
\begin{gathered}
 N_p^{\mathrm{ETAS,Omori}}=\lambar c\left[
 \frac{\theta\gamma}{\theta\gamma-1}
 \left(\frac{A}{1-p}\right)^{1/(\theta-1)}-1\right],
 \\ \qquad p_*\le p<1.
 \label{eq:app_omori_highp}
\end{gathered}
\end{equation}
The two branches are continuous at \(p_*\). Fertility heterogeneity controls the low-\(p\) shape
and the high-\(p\) amplitude, while temporal memory fixes the high-\(p\) exponent.

\section{Long-Memory Finite-Lookback Concentration}
\label{app:lookback}

The infinite-forecast weight is not normalizable over an infinite stationary past when
\(0<\theta<1\). This appendix studies the different construction obtained by retaining a finite
past of length \(H\) and then sending \(H\to\infty\). It supplies the full parametric \(f_p\)
curve, the \(\theta\gamma=1\) crossover, and both endpoint asymptotics referenced in the main
text. These are finite-lookback results, not stationarity conditions for the finite-horizon
observable.

For \(0<\theta<1\), the infinite-forecast total weight is not normalizable over an infinite
stationary past. A distinct diagnostic is obtained by declaring a lookback \(H\), setting
\(R=1+H/c\), and taking \(H\to\infty\). For \(q<\kappa_0\), write the crossover age as
\begin{equation}
 x_q=\left(\frac{\kappa_0}{q}\right)^{1/\theta}=zR,
 \qquad 0<z<1.
\end{equation}
The limiting retained weight fraction and selected catalog fraction are, for
\(\theta\gamma\ne1\),
\begin{align}
 p(z)&=\frac{(1-\theta)z^{\theta(\gamma-1)}
 -\theta(\gamma-1)z^{1-\theta}}{1-\theta\gamma},
 \label{eq:app_pz}\\
 f(z)&=\frac{z^{\theta\gamma}-\theta\gamma z}{1-\theta\gamma}.
 \label{eq:app_fz}
\end{align}
On the marginal line \(\theta\gamma=1\), the continuous limits are
\begin{equation}
\begin{gathered}
 p(z)=z^{1-\theta}\left[1+(1-\theta)\log\frac{1}{z}\right],
 \\ \qquad
 f(z)=z\left(1+\log\frac{1}{z}\right).
 \label{eq:app_marginal}
\end{gathered}
\end{equation}
Equations~\eqref{eq:app_pz}--\eqref{eq:app_marginal} parametrically define the complete limiting
curve \(f_p\).

As \(p\downarrow0\), three regimes follow:
\begin{equation}
 f_p\sim
 \begin{cases}
 \displaystyle
 \frac{1}{1-\theta\gamma}
 \left(\frac{1-\theta\gamma}{1-\theta}\right)^{\gamma/(\gamma-1)}
 p^{\gamma/(\gamma-1)}, & \theta\gamma<1,\\[9pt]
 \displaystyle
 \frac{1}{1-\theta}p^{1/(1-\theta)}
 \left(\log\frac{1}{p}\right)^{-\theta/(1-\theta)}, & \theta\gamma=1,\\[9pt]
 \displaystyle
 \frac{\theta\gamma}{\theta\gamma-1}
 \left[\frac{\theta\gamma-1}{\theta(\gamma-1)}p\right]^{1/(1-\theta)},
 & \theta\gamma>1.
 \end{cases}
 \label{eq:app_smallp}
\end{equation}
Below the crossover, the exponent is controlled by the fertility tail; above it, the exponent is
controlled by Omori memory. At \(p\uparrow1\), both sides share
\begin{equation}
 f_p=1-\frac{\gamma}{(\gamma-1)(1-\theta)}(1-p)+o(1-p).
 \label{eq:app_largep}
\end{equation}
The line \(\theta\gamma=1\) belongs to this finite-lookback, infinite-forecast construction. It is
neither the subcritical stationarity condition nor a sharp transition of the finite-horizon
observable. Continuum simulations up to one million events reproduce the memory-dominated and
marginal exponents; the fertility-dominated limit converges slowly because the selected set can
contain only a few extreme events.

\begin{figure*}[!tp]
\centering
\includegraphics[width=\textwidth,height=0.66\textheight,keepaspectratio]{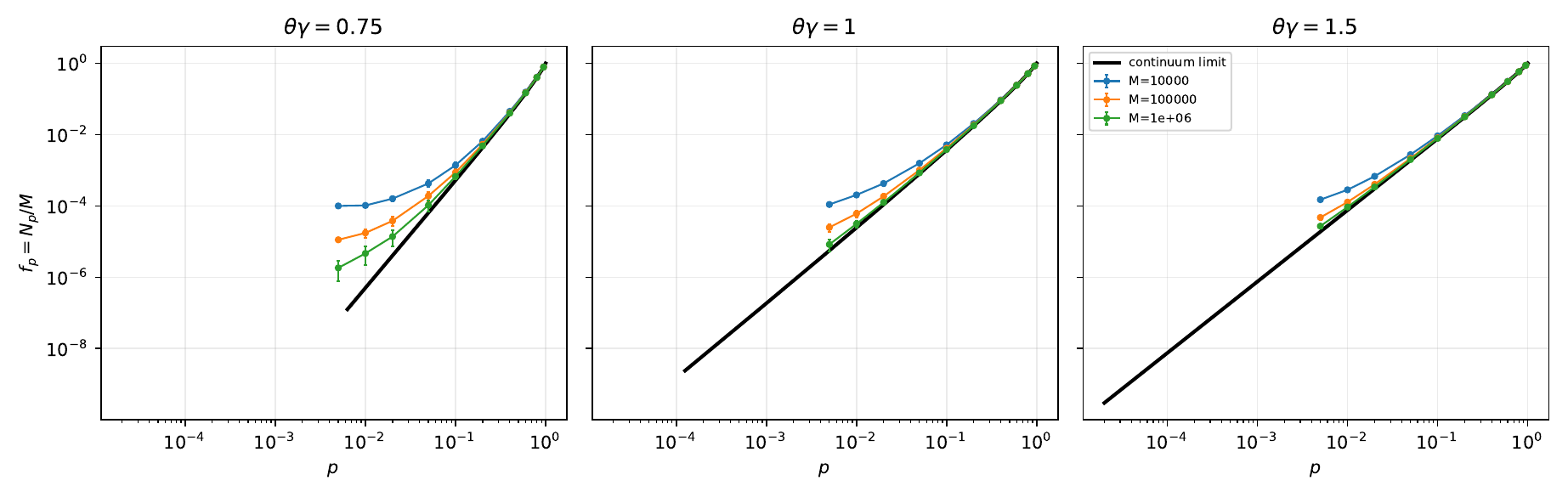}
\caption{Finite-size convergence of the finite-lookback concentration curve at fixed
\(\theta=.5\), below, on, and above the crossover \(\theta\gamma=1\). Points and bars give Monte
Carlo means and standard errors for finite populations; black curves are
equations~\eqref{eq:app_pz}--\eqref{eq:app_marginal}. The discreteness floor \(N_p\ge1\) delays the
small-\(p\) fertility-dominated asymptote even at \(10^6\) events.}
\label{fig:lookback_convergence}
\end{figure*}

Figure~\ref{fig:lookback_convergence} verifies the complete parametric curve on both sides of
the crossover and at the marginal line. Its principal numerical message is that the
fertility-dominated small-\(p\) asymptote suffers the strongest discreteness effects, whereas the
memory-dominated and marginal regimes approach their continuum predictions more directly.

\section{Exponential Scale-Invariant Weight Cloud}
\label{app:exponential_cloud}

This appendix derives the complete mapped weight process for exponential memory. The result
explains the horizon invariance of all normalized exponential concentration measures and gives
the scale-invariant small-weight cloud underlying the logarithmic high-\(p\) law. It also records
the generalized Dickman representation of single-event condensation and the Pareto-mark
extension used in the distributional checks.

For exponential memory and constant fertility, set
\begin{equation}
 w_0=n(1-e^{-\omega T}),\qquad \eta=\frac{\lambar}{\omega}.
\end{equation}
The mapped Poisson weight process has
\begin{equation}
\begin{gathered}
 r_T(w)=\frac{\eta}{w}\bm{1}_{(0,w_0)}(w),
 \\ \qquad \Lambda_T(q)=\eta\log\frac{w_0}{q},\quad 0<q<w_0.
 \label{eq:app_scale_invariant}
\end{gathered}
\end{equation}
Its total-weight transform is
\begin{equation}
\begin{gathered}
 \E[e^{-sW_T}]=\exp[-\eta\operatorname{Ein}(sw_0)],
 \\ \qquad
 \operatorname{Ein}(x)=\int_0^x\frac{1-e^{-u}}{u}du.
 \label{eq:app_dickman_transform}
\end{gathered}
\end{equation}
After conditioning on the largest weight and rescaling the smaller weights by it, their sum has a
generalized Dickman law \(D_\eta\). Consequently,
\begin{equation}
 \Prob(N_{p,T}=1)=\Prob\left(D_\eta\le\frac{1-p}{p}\right).
 \label{eq:app_dickman_condensation}
\end{equation}

For Pareto fertilities, define \(w_*=\kappa_0(1-e^{-\omega T})\). The exact L\'evy density and
tail measure are
\begin{align}
 r_T(w)&=
 \begin{cases}
 \eta/w,&0<w<w_*,\\
 \eta w_*^\gamma w^{-1-\gamma},&w\ge w_*,
 \end{cases}
 \label{eq:app_pareto_levy}\\
 \Lambda_T(q)&=
 \begin{cases}
 \eta[\log(w_*/q)+1/\gamma],&0<q<w_*,\\
 (\eta/\gamma)(w_*/q)^\gamma,&q\ge w_*.
 \end{cases}
 \label{eq:app_pareto_tail}
\end{align}
The mean total weight is finite for \(\gamma>1\), whereas
\(\operatorname{Var}(W_T)=\infty\) for \(1<\gamma\le2\). Its large tail is governed by a
single large weight, \(\Prob(W_T>x)\sim\Lambda_T(x)\propto x^{-\gamma}\).

\section{Occupation Uncertainty and Block Self-Averaging}
\label{app:occupation_variance}

This appendix provides the uncertainty calculation behind the occupation-law experiment in
Section~6. It relates the finite-record error of a calendar-time empirical CDF to the temporal
autocovariance of a bounded indicator of the triggering core. The resulting integrated
correlation time defines the effective number of independent observations and motivates the
blockwise diagnostic used to distinguish slow self-averaging from a persistent Palm/calendar
difference.

For fixed \(k\), define the bounded stationary indicator
\(X_k(t)=\bm{1}_{\{N_{p,T}(t)\le k\}}\) and its covariance
\begin{equation}
 C_k(s)=\operatorname{Cov}[X_k(0),X_k(s)].
\end{equation}
The exact variance of the length-\(L\) occupation CDF is
\begin{equation}
 \operatorname{Var}[\widehat F_{p,T;L}^{\mathrm{occ}}(k)]
 =\frac{2}{L^2}\int_0^L(L-s)C_k(s)ds.
 \label{eq:app_occ_variance}
\end{equation}
If
\begin{equation}
 \tau_k^{\mathrm{int}}=\frac{1}{C_k(0)}\int_0^\infty C_k(s)ds<\infty,
\end{equation}
then
\begin{equation}
 \operatorname{Var}[\widehat F_{p,T;L}^{\mathrm{occ}}(k)]
 \sim\frac{2\tau_k^{\mathrm{int}}}{L}
 F_{p,T}^{\mathrm{ann}}(k)[1-F_{p,T}^{\mathrm{ann}}(k)].
 \label{eq:app_occ_asymptotic}
\end{equation}
Thus the effective number of independent calendar observations is of order
\(L/(2\tau_k^{\mathrm{int}})\), rather than the number of grid evaluations.

For a block \(b\) of duration \(L\), let \(\widehat F_b(k)\) be its occupation CDF. A direct
ergodicity diagnostic is
\begin{equation}
\begin{gathered}
 E_L(k)=\frac{1}{B}\sum_{b=1}^{B}
 [\widehat F_b(k)-F_{p,T}^{\mathrm{ann}}(k)]^2,
 \\ \qquad
 E_L=\sum_{k\ge1}\omega_kE_L(k),
 \label{eq:app_block_error}
\end{gathered}
\end{equation}
with nonnegative weights summing to one. Under a mixing stationary null these discrepancies
decay with \(L\), at a rate governed by the same serial dependence. Event-time sampling instead
converges to a Palm law and need not approach the calendar-time CDF.

\section*{Acknowledgments}
The authors thank the institutions supporting this work. Grant identifiers and individual
acknowledgments will be added before submission.

\section*{Open Research Statement}
The CAT5 catalog is openly available from Zenodo
\cite{Tan2021AmatriceDataset}.

\bibliography{references}

\end{document}